\documentclass{mn2e}
\usepackage{times}
\usepackage{epsfig}
\usepackage{amsmath}
\usepackage{amsfonts}
\usepackage{textcomp}
\usepackage{longtable}

\setlongtables

\title[]{2-D Monte Carlo simulations of H~{\sc i} line formation in massive YSO disk winds}
\author[S. A. Sim, J. E. Drew, K. S. Long]{S. A. Sim\thanks{s.sim@imperial.ac.uk}$^1$, J. E. Drew$^1$, K. S. Long$^2$ \\
$^1$Astrophysics Group, Imperial College London,
Blackett Laboratory, Prince Consort Road, London, SW7 2AZ, UK\\
$^2$Space Telescope Science Institute, 3700 San Martin Drive, Baltimore, MD 21218, USA}

\date{3 August 2005}

\pagerange{\pageref{firstpage}--\pageref{lastpage}}
\pubyear{2005}
\begin{document}
\maketitle
\label{firstpage}

\begin{abstract}
Massive young stellar objects (YSOs) are powerful infrared H~{\sc i} line emitters.
It has been suggested that these lines form in a outflow from
a disk surrounding the YSO. Here, new two-dimensional Monte Carlo radiative transfer 
calculations are described which test this hypothesis. 
Infrared spectra are synthesised for a YSO disk wind model based on earlier hydrodynamical
calculations. The model spectra are in qualitative agreement with
the observed spectra from massive YSOs, and therefore provide support for a disk wind explanation
for the H~{\sc i} lines. However, there are some significant differences:
the models tend to overpredict the Br$\alpha$/Br$\gamma$ ratio of equivalent-widths and produce
line profiles which are slightly too broad and, in contrast to typical observations, are 
double-peaked. 
The interpretation of these differences within the context of the disk wind picture 
and suggestions for their resolution via modifications to the assumed disk and outflow structure are discussed. 
\end{abstract}


\section{Introduction}

Understanding the formation and pre-main-sequence evolution of massive stars ($M \ga 10$~M$_{\odot}$) 
is an important goal in modern astrophysics. 
Currently, the formation of low-mass stars 
via gravitational collapse of a molecular cloud and subsequent disk accretion is, at least 
conceptually, well understood but it is not clear whether massive stars form through a scaled-up
version of the same process or whether more complex, environmental effects are important (see e.g.
Bonnell, Vine \& Bate 2004 for a recent discussion). 

Since massive stars form inside dense molecular clouds, they are observed primarily at infrared (IR)
and longer wavelengths.
Embedded massive YSOs are powerful sources of IR H~{\sc i}
line emission (Simon et al. 1981, 1983; Drew, Bunn \& Hoare 1993; Bunn, Hoare \& Drew 1995; Blum et al. 2004)
with equivalent widths (EWs) in Br$\alpha$ of $\sim 5$ -- 90~\AA~(Bunn et al. 1995). 
The lines profiles are complex, displaying both fairly narrow 
(full-width-at-half-maximum [FWHM] $\sim$ 50 -- 100~km~s$^{-1}$) line cores
and broad wings extending out to $\sim 400$~km~s$^{-1}$ in extreme cases. 
In the sample of objects observed by Bunn et al. (1995), the profiles were always 
single-peaked but recent observations reported by Blum et al. (2004)
include at least one object (NGC~3576) which shows clear evidence of double-peaked profiles.  
By examining flux ratios across the profiles of lines with differing opacities, Bunn et al. (1995) found
evidence that, at least in some sources, the lines form in an accelerating outflow. 

Normal, main-sequence OB stars also show H~{\sc i} emission lines. These are produced in the star's 
fast ($\sim 1000$~km~s$^{-1}$)
spherical wind, but the strength of emission
in YSO spectra is much greater than in O-stars. If the features in YSOs were due to reprocessing of radiation
in a spherically symmetric wind, the implied mass-loss rates would be up to $10^{-6}$~M$_{\odot}$~yr$^{-1}$ 
(Simon et al. 1983),
substantially exceeding those of field stars of comparable spectral type.
Therefore, to interpret the H~{\sc i} observations it is worthwhile to consider alternatives
to formation in a spherically symmetric structure.

Important observational constraints on the geometry of emitting material in YSOs are provided by 
the first overtone bands of CO at 2.3$\mu$m 
(e.g. Carr 1989; Carr et al. 1993; Chandler et al. 1993; Chandler, Carlstrom \&
Scoville 1995).
In particular, both
Carr et al. (1993) and Chandler et al. (1995) conclude that models based on 
an accretion disk surrounding the central object reproduce the
CO observations of YSOs.
Furthermore, although there are some cases in which CO data can be modelled in terms of a
stellar wind, in general the disk model encounters fewer difficulties (see Chandler et al. 1995).
The observations strongly suggest that the CO emission originates close to the
central star: Bik \& Thi (2004) and Blum et al. (2004) 
have recently studied CO emission
from a range of massive YSOs and derived disk radii in the range 0.1 -- 5~AU.
Thus it seems probable that YSOs harbour accretion disks containing significant amounts of
hot gas fairly close to the central object and it is natural to consider whether the H~{\sc i} emission
is associated with such a disk or its environment.

Hamann \& Simon (1986) suggested that powerful recombination line emission of the sort discussed
above originates in an outflow associated not with a stellar wind but with mass-loss
from a disk. Their work focused on MWC~349, an object whose evolutionary status is unclear
but which possesses spectroscopic signatures (including H~{\sc i} and CO emission features) broadly 
similar to the embedded massive YSOs (see Hamann \& Simon 1986; Kraus et al. 2000). 
Hollenbach  et al. (1994) developed a model for outflow from young massive stars by photoevaporation
of the outer regions of their disks. They were motivated by the need to explain the high frequency of
occurrence of ultracompact H~{\sc ii} regions -- photoevaporation can provide a source of mass input
for H~{\sc ii} regions to balance the loss by pressure-driven expansion. 
Although the rate of mass-loss in photoevaporation models is large, it 
occurs at large radii (typically $> 100$~AU) and the characteristic flow velocities are low 
(comparable to the typical sound speed of 10 -- 50 km~s$^{-1}$, Hollenbach et al. 1994). Thus, 
such a flow does not provide a promising origin 
for relatively high velocity features, such as the IR H~{\sc i} broad line wings (Bunn et al. 1995).

Recently,
Drew, Proga \& Stone (1998) proposed that the intense radiation field 
produced by a massive YSO may drive mass-loss 
from the surface of the inner parts of its disk. They performed a hydrodynamical
simulation which showed that, in addition to driving a normal hot star 
wind component, 
radiation pressure from a massive YSO could propel
a dense equatorial flow from a surrounding disk with terminal velocity in the same
regime as the observed H~{\sc i} linewidths. These results led
them to speculate that 
such a model accounts for the IR H~{\sc i} lines. However, they did not perform
the radiative transfer calculations required to confirm this conjecture.

To investigate the possibility that massive YSOs harbour disk winds which give rise to the
observed IR H~{\sc i} line emission, this paper presents the results of new two-dimensional
radiative transfer calculations which account for both the complex geometry of a YSO disk
wind (starting from, but not limited to, the Drew et al. 1998 model) and the detailed atomic physics of  
H~{\sc i} line formation.

In Section 2, the YSO model adopted in this investigation and its parameters are 
discussed. The radiative transfer calculations have been performed using the Monte Carlo (MC)
code described by Long \& Knigge (2002) after incorporating a sophisticated
treatment of H~{\sc i} line formation using the approach described by
Lucy (2002,2003); the method of 
calculation and code used are discussed in Section 3. 
Results are presented in Section 4 and our conclusions discussed in Section 5.

\section{Model}

For the radiative transfer calculations (Section 4) a simply parameterised model for the
YSO and its disk is adopted. 
In choosing the parameters for this model, our starting point has been
to construct a reasonable representation of the 
disk wind obtained in the hydrodynamical simulations presented by Drew et al. (1998). Therefore, 
before discussing our model in detail, a brief description of this hydrodynamical model is given.

\subsection{Summary of the hydrodynamical disk wind model}

Following the methods of Proga et al. (1998),
Drew et al. (1998) performed a hydrodynamical simulation of 
a radiatively driven wind from a disk around a massive YSO.
They considered an early-B type star (mass $M_{*} = 10$~M$_{\odot}$,
luminosity $L_{*} = 8500$~L$_{\odot}$, radius $r_{*} = 5.5$~R$_{\odot}$)
surrounded by an accretion disk extending from the stellar surface to 
an outer radius $r_{\mbox{\scriptsize disk}} = 10$~$r_{*}$, the outer boundary of their
computational domain. Noting that,
for reasonable accretion rates ($\sim 10^{-6}$~M$_{\odot}$~yr$^{-1}$),  
the luminosity of the central star far exceeds the accretion luminosity of the
disk, they accounted for 
reprocessing of stellar light by the disk and then computed
the radiation force using the Castor, Abbott \& Klein (1975) 
parameterisation of the force due to spectral lines.

Their simulation exhibited a steady state solution with two fairly distinct 
outflow components: 
a fast ($\sim 2000$~km~s$^{-1}$) polar wind from the central star, 
rather similar to a normal 
hot star wind; and a slower ($< 400$~km~s$^{-1}$), denser equatorial 
outflow from the disk 
at colatitude $\theta \sim 60$\textdegree~ with mass-loss rate 
$\sim 3 \times 10^{-8}$~M$_{\odot}$~yr$^{-1}$.
Between these two outflow components they found a complex transitional zone 
in which the stellar
wind streamlines were compressed due to 
non-radial 
radiation force components and the presence of the disk wind.

Drew et al. (1998) suggested that the high density and moderate velocity 
of the equatorial disk-wind 
component make it a promising site for the formation of the H~{\sc i} 
lines observed in massive
YSO spectra -- our goal is to present radiative transfer calculations 
to examine this hypothesis.

\subsection{Geometry}

Figure~1 illustrates 
the geometrical construction used to
describe the wind, parameters for
which are discussed in the next sub-section.
In order to capture the essence of the Drew et al. (1998) hydrodynamical 
disk wind model
(described above),
the structure consists of the 
following four
basic elements:

\begin{enumerate}

\item
A central star. The star is assumed to be spherical with radius $r_{*}$. 

\item
An accretion disk.  
The disk lies in the $xy$-plane of the
adopted coordinate system and 
is assumed to extend from the surface
of the star to an outer radius $r_{\mbox{\scriptsize disk}}$. 
For simplicity, it is assumed to be geometrically thin and flat.

\item
A disk wind. The wind, which is launched from the disk, is described following 
Long \& Knigge (2003) and Knigge, Woods \& Drew (1995): namely the streamlines of the 
disk wind are assumed to converge at a point which lies 
on the $z$-axis 
a distance $d$ below the 
coordinate-origin. 
Thus the angular boundaries of the disk wind are 
determined by the choice of the inner and 
outer disk radii and the distance $d$; namely
$\theta_{\mbox{\scriptsize min}} = \tan^{-1} (r_{*} / d)$
and 
$\theta_{\mbox{\scriptsize max}} = \tan^{-1} (r_{\mbox{\scriptsize disk}} / d)$(see Figure~1).
Motivated by the Drew et al. (1998) simulation, the flow is assumed to be stationary and smooth.

\item
A spherical stellar wind component. This is assumed to occupy the entire polar region
above the disk wind and to have radial streamlines. The
boundary between the stellar wind and disk wind is not treated in detail; it is
assumed to be perfectly sharp -- thus the transition zone between the disk and stellar
winds which exists in the Drew et al. (1998) simulation is neglected here.

\end{enumerate}

\begin{figure}
\epsfig{file=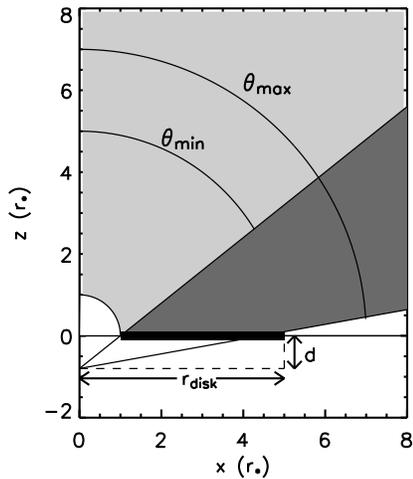, height=7cm}
\caption{
The elements of the
geometrical construction used to define the wind (only the positive $xz$-plane 
is shown -- the wind is symmetric about both the $xy$-plane and the $z$-axis).
The quarter-circle line around the origin represents the stellar surface;
the stellar wind is shaded light grey, the disk wind dark grey and the disk 
black. The equatorial region beyond the disk (white) is assumed to be empty.
The figure is {\it not to scale} for the parameters given in Table~1.
}
\end{figure}

\subsection{Model parameters}

Table~1 gives 
a list of the wind parameters adopted for the reference model 
(hereafter, Model~A) which are discussed individually below.
Of our models, Model~A and the closely related Model~B (see Section 4.3) 
are those most closely matching the Drew et al. (1998) simulation.

\subsubsection{Central star}

The parameters of the central star are 
those of a B1-B2 main sequence star.
The effective temperature was computed from the luminosity 
using the Stefan-Boltzmann
equation. The star is assumed to emit radiation as a black body.

\subsubsection{Accretion disk}

In our reference model, $r_{\mbox{\scriptsize disk}} = 10$~$r_{*}$ is adopted --
the same computational domain considered by Drew et al. (1998).
Realistically, the disks around massive YSOs are likely to extend to significantly
larger radii. Therefore, we will also consider disks extending to 
$r_{\mbox{\scriptsize disk}} = 100$~$r_{*}$ in Section~4. To consider even larger
radii becomes impractical since increasing the physical size of the calculation 
limits the spatial resolution available in the inner regions -- computing
feasibility prevents us from having arbitrarily many grid cells. It is noted that,
at least for the treatment of disk radiation adopted in our reference model, 
considering an even larger outer radius will have negligible effect -- 
the disk temperature at such large radii will be too low for its thermal radiation to 
contribute significantly to the IR spectral range under consideration.

\begin{figure}
\epsfig{file=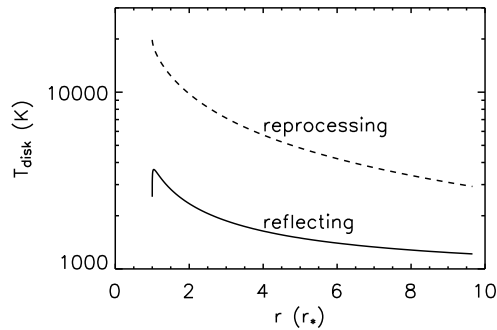, height=5cm}
\caption{The disk temperature ($T_{\mbox{\scriptsize disk}}$)
as a function of radius ($r$) for a reflecting disk (solid line)
and a reprocessing disk (dashed line). The reflecting disk 
temperatures are always lower
since they only describe the accretion luminosity while
the reprocessing disk also accounts for re-radiation of energy absorbed from the star.
}
\end{figure}

Given that it lies considerably beyond the scope of this investigation to study 
radiative transfer within the accretion disk in detail, approximations must be made
regarding the treatment of the disk and its emission. 
In the calculations presented in Section~4, results using two different approaches
are considered. It is known from observations (Henning, Pfau \& Altenhoff 1990) that
there are significant differences in the IR SEDs of massive YSOs -- some have large
near-IR excesses while other have little or no excess.
Therefore, we have chosen our two models for 
disk emission to approximately bracket the range of plausible IR SEDs
for massive YSOs.

\begin{figure*}
\epsfig{file=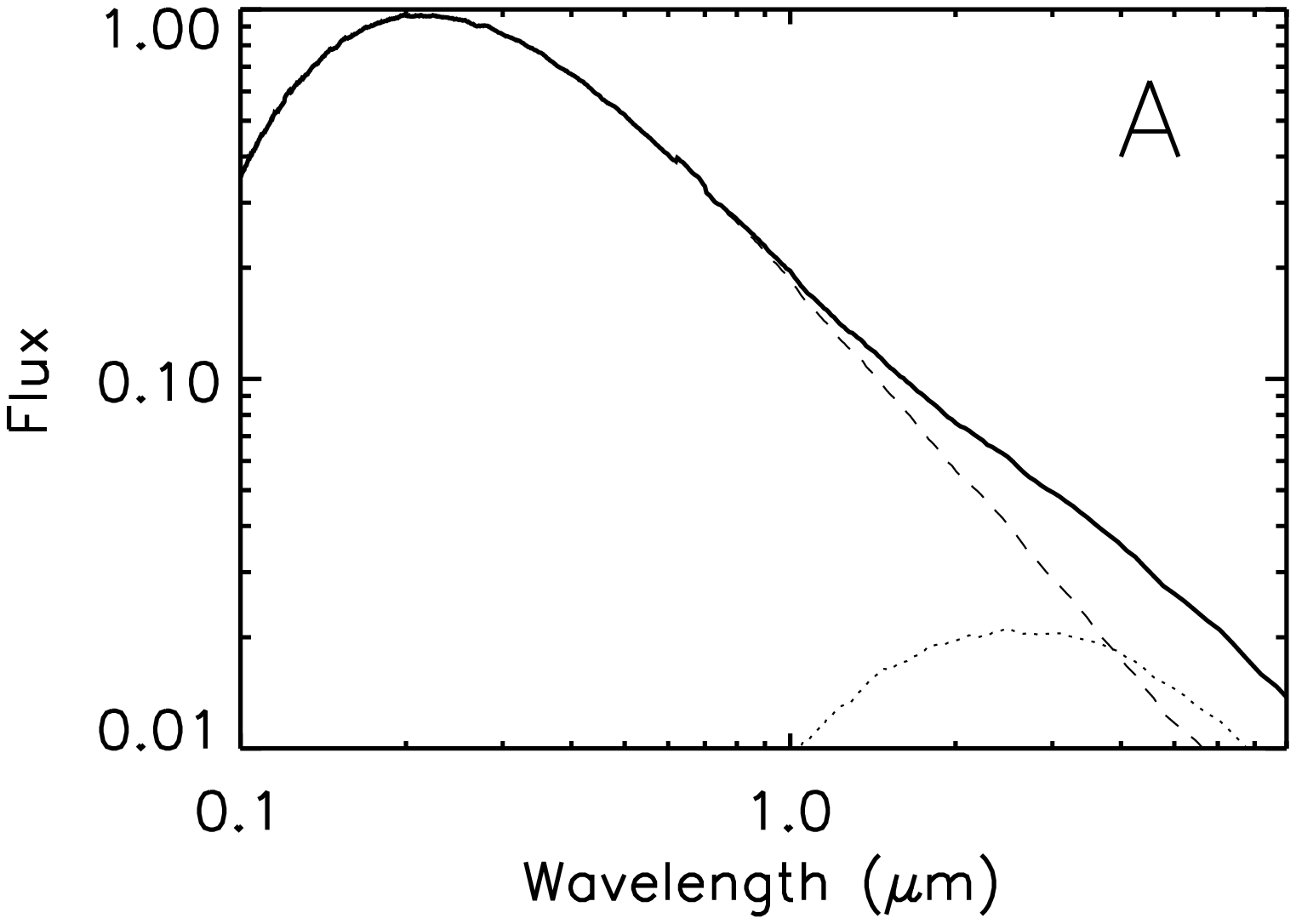, height=4cm}
\epsfig{file=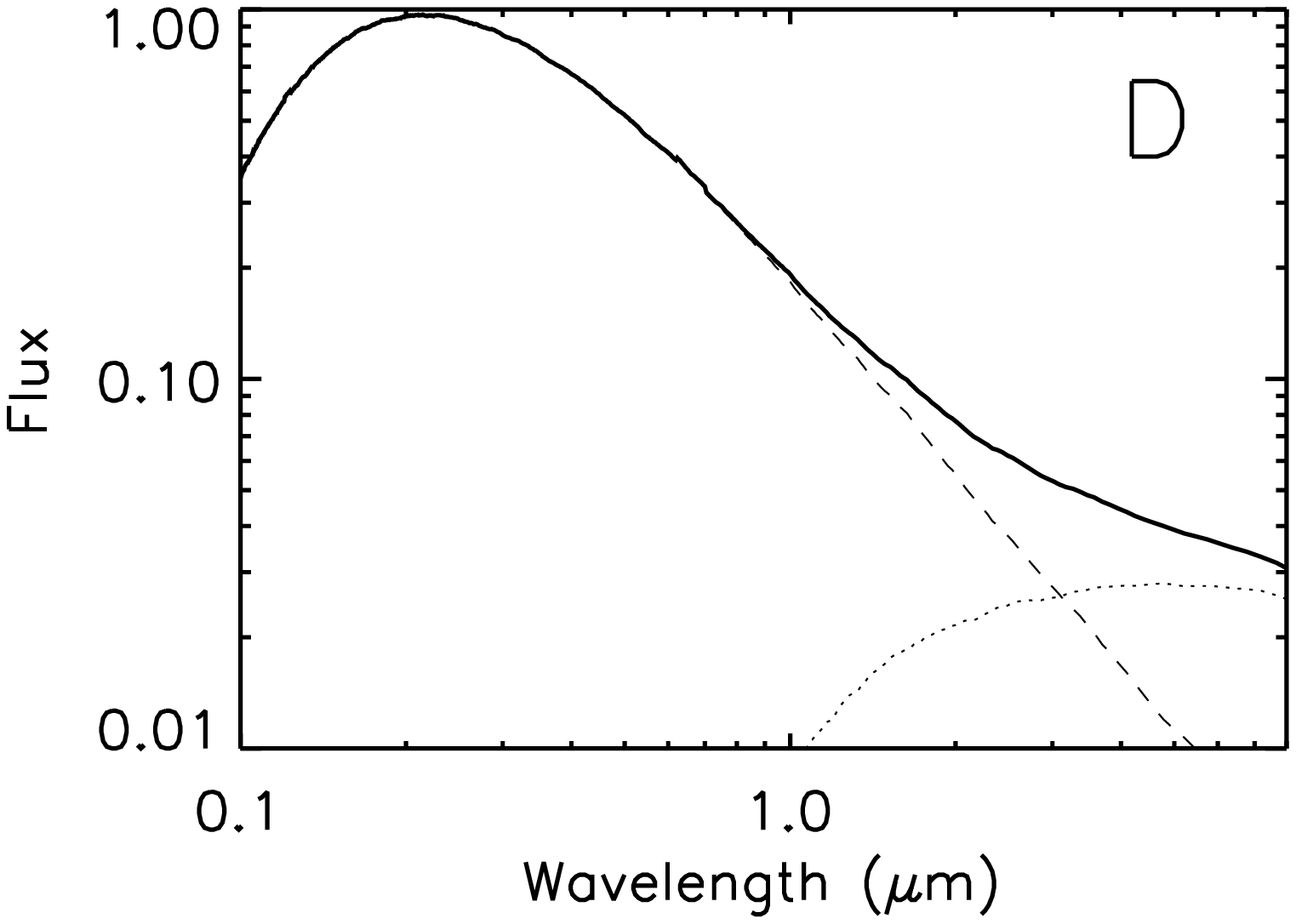, height=4cm}\\
\epsfig{file=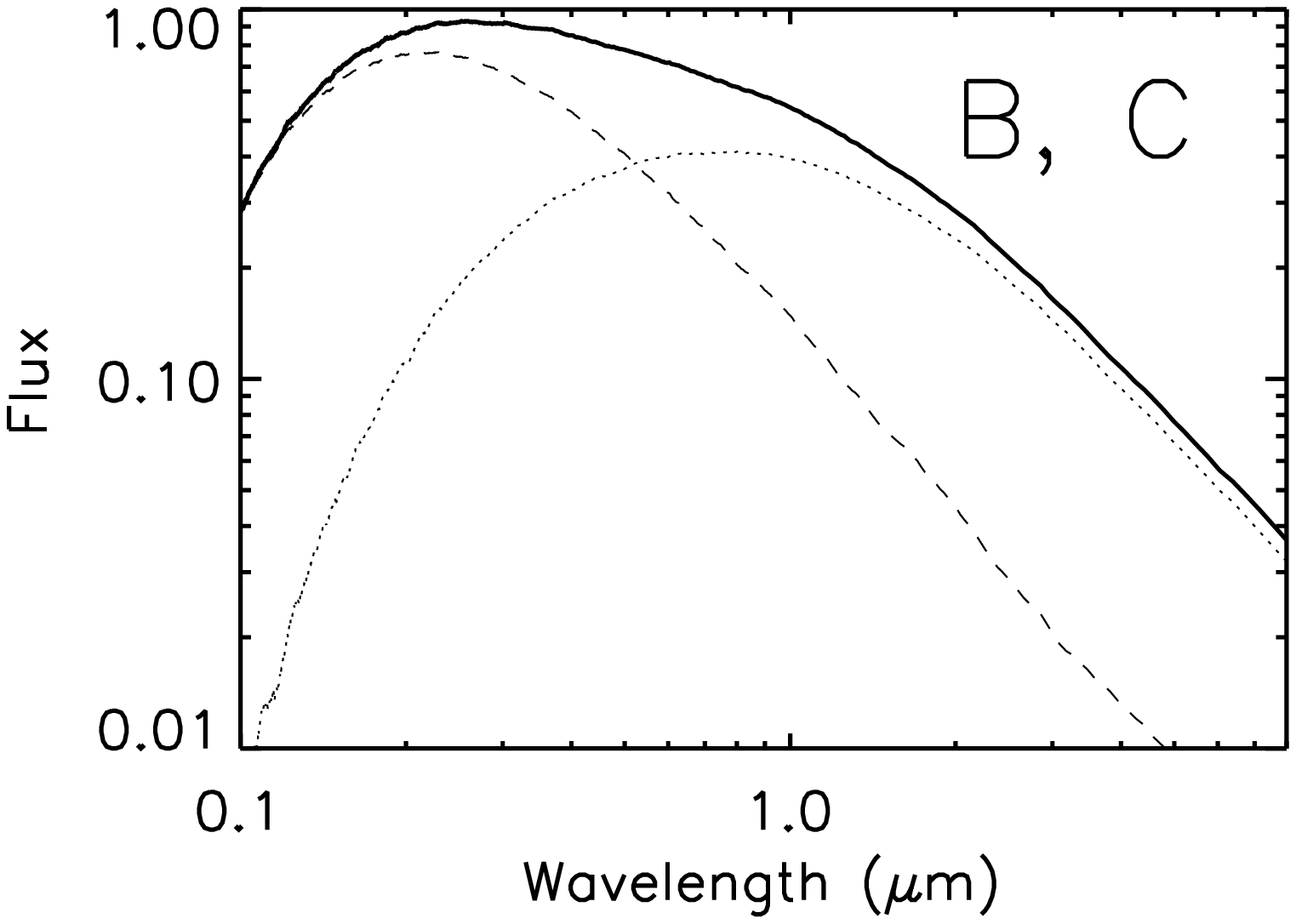, height=4cm}
\epsfig{file=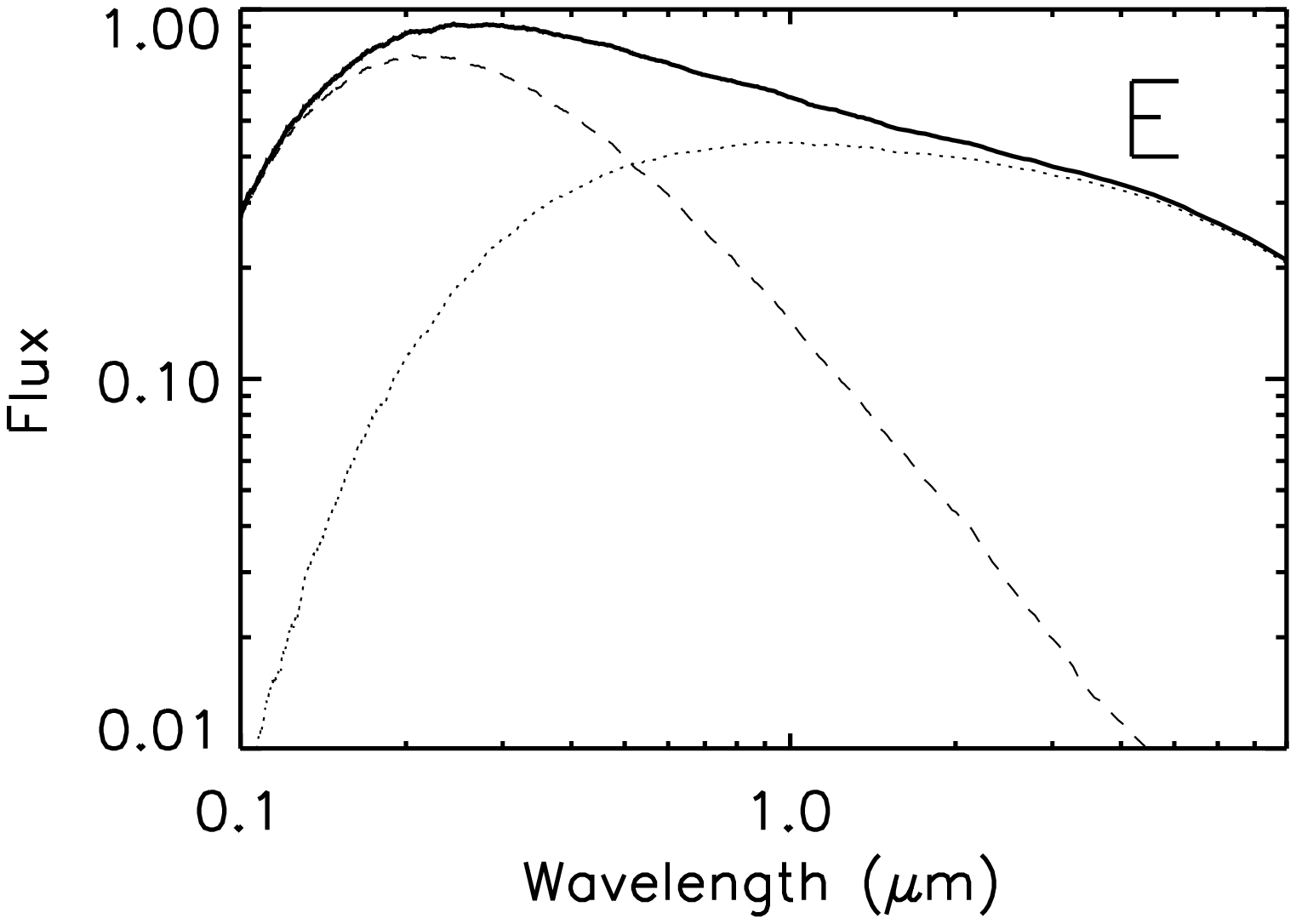, height=4cm}\\
\caption{Spectral energy distributions (SED) for models with star and reflecting disk
(upper panels) and star and reprocessing disk (lower panels). The left panels
are for disks extending to outer radius $r_{\mbox{\scriptsize disk}} = 10$~$r_{*}$ while
the right panels are appropriate for $r_{\mbox{\scriptsize disk}} = 100$~$r_{*}$.
In each panel, the heavy line shows the complete SED while the dashed and dotted lines
show, respectively, the contributions due to the star and disk. The same relative
flux scale is used in each plot. The letters given in the upper right corners of the panels 
indicate in which of the models discussed in Section~4 the SED is adopted.}
\end{figure*}

In some cases, referred to as a ``reflecting'' disk, it is assumed that the disk 
effective temperature is determined as a function of radius by the accretion rate
(exactly as discussed by Long \& Knigge 2002) and that annuli of the disk
emit as black bodies at the local temperature. During the MC 
simulations, it is assumed that any radiation which strikes the disk from above is
reflected by a hot disk atmosphere rather than absorbed and reprocessed. 
Since the computational domain is symmetric about the disk plane, this is
closely equivalent to the assumption that the disk is completely optically thin (transparent).
In Model~A,
the disk is assumed to be ``reflecting''.

Alternatively, as a simple ``reprocessing'' disk model, it is assumed that stellar radiation 
falling on the disk is absorbed and the energy re-radiated.
In this case, the disk is divided into concentric 
annuli and the energy falling on each annulus from the star computed following
the discussion of Proga et al. (1999). By assuming that all this energy is
re-radiated locally, and that the annuli radiate as black bodies, the
disk temperatures are then obtained. The accretion luminosity is still included
in this model but is small relative to the 
reprocessed luminosity. At the start of the MC simulation, the disk annuli emit black-body radiation 
according to their effective temperatures -- thereby accounting for the luminosity 
due to reprocessing of stellar light --  
and during the subsequent propagation of MC quanta,
any radiation which strikes the disk is simply absorbed.

Computed disk temperature for both reflecting and reprocessing disks are shown, 
as a function of radius, in Figure~2. Spectral energy distributions (SED) for radiant energy 
emitted from models with both reflecting and reprocessing disks are shown in Figure~3.
Comparing the SEDs for reflecting and reprocessing disks with $r_{\mbox{\scriptsize disk}} = 10$~$r_{*}$
(left-hand panels in Figure~3) shows that the treatment of the disk makes a significant
difference to the continuum brightness in the IR region of interest.
If the disk only reflects (or transmits) light which strikes it,
the $\sim$ 2 -- 4~$\mu$m
continuum is a combination of
stellar radiation and accretion luminosity released in the disk.
(Figure~3, upper-left panel).
However, if the 
disk thermally reprocesses all the
stellar light which strikes it, the reprocessed light is dominant (lower-left panel).
Increasing the outer radius of the disk to 100~$r_{*}$ increases its brightness. With a reflecting disk,
this effect is fairly small below about 4~$\mu$m (upper-right panel) because the temperatures of the outer
parts of the disk are very low. However, since the reprocessing disk has higher temperatures (Figure~2),
changing its outer radius has a greater effect in the waveband of interest (Figure~3, lower-right panel).

We note that since the treatment of the radiation force is grey in the Drew et al. (1998) simulation and that
in both our approaches to the disk SED all the radiative flux which falls on the disk is re-emitted locally, both 
our disk SED models are equally valid in the context of the hydrodynamical calculations. 

\subsubsection{Disk wind}

The geometry of the disk wind is specified by the distance between the
origin and the focus point, $d$ (see Figure~1). Here, $d = 0.14$~$r_{*}$
is adopted
leading to $\theta_{\mbox{\scriptsize min}} = 82$\textdegree~and 
$\theta_{\mbox{\scriptsize max}} = 89$\textdegree~(see above).
The dense disk wind component present in the hydrodynamic model of
Drew et al. (1998) occupied a rather wider angular range ($\sim 25$\textdegree).
However, Proga et al. (1999) showed that improved treatment of the 
radiative line force leads to significantly more swept back equatorial disk winds.
In particular, for a model in which the luminosity of the central object
substantially exceeds that of the disk (their model E) they found a disk
wind with opening angle 8\textdegree. Thus the $d$-value adopted here
is chosen to reflect the narrow disk winds obtained by Proga et al. (1999).

The density and velocity structure of the disk wind closely follows that
adopted by Long \& Knigge (2002). Following equation~(7) of Long \& Knigge 
(2002), it is assumed that the mass loss rate per unit area from the
disk can be described by

\begin{equation}
\frac{\delta \dot{M}}{\delta A} \propto 
T_{\mbox{\scriptsize disk}}(r)^{4\alpha}
\end{equation}
On the assumption that the mass-loading is proportional to the local
luminous flux, $\alpha = 1$ is adopted here. The total mass-loss rate in the disk
wind is fixed at the mass-loss rate obtained by Drew et al. (1998): 
$\dot{M} = 3 \times 10^{-8}$~M$_{\odot}$~yr$^{-1}$.
Note that the improved treatment of the line force (i.e. Proga et al. 1999)
did not significantly change the total mass-loss rates from those
of the earlier treatment.
The poloidal velocity on a streamline
in the disk wind is described using equation~(8) of
Long \& Knigge (2002), namely

\begin{equation}
v(l) = c_{s} + (v_{\infty} - c_{s}) \left( {1 - \frac{R_{d}}{l + R_{d}}} \right)^{\beta_{d}}
\end{equation}
where $c_{s}$ is the sound speed on the disk surface at the base of the 
streamline,
$v_{\infty}$ is the terminal velocity of the streamline, $l$ is the poloidal distance, 
$R_{d}$ is a velocity-law
scale length for the disk wind and $\beta_{d}$ is an exponent which 
determines the acceleration of the wind. 
For simplicity,
$v_{\infty} = v_{\mbox{\scriptsize esc}}$ is adopted where 
$v_{\mbox{\scriptsize esc}}$ is the escape velocity on the disk surface
at the base of the streamline. Figure~4 shows 
$v_{\infty}$ as a function of $\theta$ -- these values compare favourably
with the terminal velocities found by Drew et al. (1998, their figure 3).
To describe the acceleration of the wind,
a velocity scale length for the disk-wind, $R_{d} = r_{*}$
and exponent $\beta_{d} = 1.5$ are used.

\begin{figure}
\epsfig{file=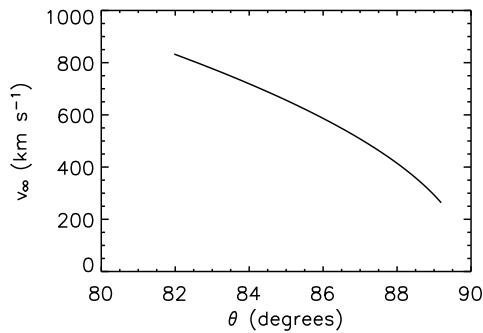, height=5cm}
\caption{The disk wind terminal velocity ($v_{\infty}$) versus
$\theta = \tan^{-1} (r/d)$ where $r$ is the disk radius at the base of
the streamline.}
\end{figure}

\subsubsection{Stellar wind}

The density of the
stellar wind component is chosen such that it would have
mass-loss rate $\dot{M}_{s} = 10^{-8}$~M$_{\odot}$~yr$^{-1}$
if it were spherically symmetric.
Following e.g. Lucy \& Abbott (1993), 
the velocity in this component is assumed to be radial and given by

\begin{equation}
v = v_{c} + (v_{\infty} - v_{c}) ( 1 - (r_{*} / r))^{\beta_{s}}
\end{equation}
Based on the Drew et al. (1998) simulations, $v_{\infty}
= 2000$~km~s$^{-1}$ is adopted. In addition, $v_{c} = 10^{-2} v_{\infty}$ is 
assumed and  
$\beta_{s} = 1$ -- a standard value in modelling hot star winds -- is chosen.

\begin{table}
\begin{center}
\caption{Parameters for the reference model (Model A). Parameters for which
departures from the reference values will be discussed are indicated with a tick~(\checkmark) 
in the third column.}
\begin{tabular}{llc} \hline
Parameter & Value & Varied\\ \hline
\multicolumn{2}{l}{Central star}\\
mass, $M_{*}$ & 10~M$_{\odot}$\\
radius, $r_{*}$ & 5.5~R$_{\odot}$\\
luminosity, $L_{*}$ & 8500~L$_{\odot}$\\
temperature, $T_{\mbox{\scriptsize eff}}$ & $2.37 \times 10^4$~K\\ \hline
\multicolumn{2}{l}{Accretion disk}\\
inner radius, $r_{\mbox{\scriptsize in}}$ & $r_{*}$ & \checkmark\\
outer radius, $r_{\mbox{\scriptsize disk}}$ & 10~$r_{*}$ & \checkmark\\
accretion rate, $\dot{M}_{\mbox{\scriptsize acc}}$ & $10^{-6}$~M$_{\odot}$~yr$^{-1}$\\
\hline
\multicolumn{2}{l}{Disk wind}\\
focus point, $d$ & 0.14~$r_{*}$& \checkmark\\
mass-loss rate, $\dot{M}$ & $3 \times 10^{-8}$~M$_{\odot}$~yr$^{-1}$& \checkmark\\
mass-loading exponent, $\alpha$ & 1\\
terminal velocity, $v_{\infty}$ & $v_{\mbox{\scriptsize esc}}$\\
acceleration length, $R_{d}$ & $r_{*}$\\
acceleration exponent, $\beta_{d}$& 1.5\\
\hline
\multicolumn{2}{l}{Stellar wind}\\
mass-loss rate, $\dot{M}_{s}$ & $10^{-8}$~M$_{\odot}$~yr$^{-1}$\\
terminal velocity, $v_{\infty}$ & 2000~km~s$^{-1}$\\
launch velocity, $v_{c}$ & $10^{-2}v_{\infty}$ \\
acceleration exponent, $\beta_{s}$& 1\\ \hline
\end{tabular}
\end{center}
\end{table}

\subsubsection{Departures from the reference model parameters}

In Section~4, the dependence of the results on several of the important
model parameters listed in Table~1 will be investigated. The calculation is, of course, particularly sensitive
to the choice of density in the wind. The density in turn is sensitive to many of the model 
parameters -- most fundamentally, the mass-loss rates which determine the mass-loading of the 
stellar and disk winds. However, the velocity-law parameters (e.g. $v_{\infty}$, $R_{d}$ and $\beta_{d}$) 
also play a significant role, as does the geometry of the disk wind (determined
by $d$). In Section~4, models with density higher than that in Model~A will be considered: these are created
by increasing the mass-loss rate only -- physically this is the parameter which leads to a simple increase
in density everywhere in the wind -- but it should be noted that there is a degree of degeneracy between
this and the other parameters mentioned above.

The second departure from the reference model (Model~A) that will be discussed in Section~4 is an increase
in the outer disk radius, $r_{\mbox{\scriptsize disk}}$. The relevance of this parameter to the determination of the disk SED has 
already been discussed (see Section~2.3.2 and Figure~3).

Finally, a model in which an inner hole is introduced to the accretion disk will be considered. In addition to
increasing $r_{\mbox{\scriptsize in}}$ for this model, the $d$-value is changed in order to preserve the same opening angles
($\theta_{\mbox{\scriptsize min}}$ and $\theta_{\mbox{\scriptsize max}}$) for the disk wind.

Departures of the other parameters from their reference values will not be considered here. 
For this investigation we wish to restrict ourselves to
central objects with parameters suitable for early-type near-main-sequence stars -- thus we
do not consider models with large values of $r_*$.
An increase in the stellar radius, $r_*$ would mimic many of the effects associated with the introduction of an 
inner hole to the disk since both push the wind out to regions where the rotational velocity is lower. Thus, 
our model with an inner hole can be 
regarded as a proxy for some of the most important consequences of an enlarged central star.

An increase 
in the stellar luminosity would clearly affect the SED, but could also change the mass-loss
rate and possibly the wind geometry -- further hydrodynamical calculations would be required to fully investigate
results for a range of luminosities. The mass-accretion rate ($\dot{M}_{\mbox{\scriptsize acc}}$) is relevant only
to calculations with reflecting disks -- in such cases, $\dot{M}_{\mbox{\scriptsize acc}}$ plays an important role in determining
the disk temperature and IR SED. For reprocessing disks, these are controlled by the
stellar radiation field and $\dot{M}_{\mbox{\scriptsize acc}}$ plays only a minor role. It is very difficult to reliably
obtain $\dot{M}_{\mbox{\scriptsize acc}}$ by observation or theory and so we restrict ourselves to 
considering only one value, that used by Drew et al. (1998).

\subsection{Atomic data}

For simplicity, it is assumed that both the disk wind and stellar wind
consist entirely of hydrogen. 
An atomic model with energy levels for principle
quantum number $n=1$ to 20 and the continuum state is used in the 
calculations. Bound-bound oscillator strengths are taken from
Menzel \& Pekeris (1935). Bound-bound collision rates are derived
from the oscillator strengths using the van Regemorter (1962) formula.
Photoionization rates are taken from TOPBASE (Cunto et al. 1993) and
collisional ionization rates are computed using equation 5-79 from 
Mihalas (1978). 
In addition to these processes associated with atomic hydrogen,
free-free absorption and emission by H~{\sc ii} ions and 
Thompson scattering by free-electrons are included in the calculations.
All other processes are neglected.

\section{Method of calculation}

In the disk wind model, the IR H~{\sc i} line formation is expected to be
driven by recombination of ionised
hydrogen gas fairly close to the central star. The observed line ratios (e.g. Bunn el al. 
1995) indicate, however, that the 
line emission is not optically thin making it necessary to perform detailed non-LTE 
radiative transfer calculations to reliably model the line formation process.

The radiative transfer calculations discussed in the next section were performed
using a modified version of the MC code written by 
Long \& Knigge (2002). 
Since the code has been described in detail elsewhere, 
only a brief overview and 
discussion of the important modifications made for this investigation 
are presented here
(see Long \& Knigge 2002 for further details of the code).

To synthesise spectra, the code performs a sequence of MC radiative
transfer calculations. The code uses indivisible energy packets as the
elementary MC quanta, assumes radiative, statistical and thermal 
equilibrium in the wind, and utilises a Sobolev treatment of bound-bound 
transitions.

The code presented by Long \& Knigge (2002) adopted a two-level approximation
in the treatment of line scattering and used approximate excitation and
ionization formulae for the computation of level emissivities. These
approximations made the code efficient and thus able to handle a large 
set of atomic data for many chemical elements. However, the two-level
approximation also meant 
that it was not suitable for
modelling lines formed by non-resonance scattering or recombination.

Since the IR H~{\sc i} lines form primarily by recombination,
the code has been substantially modified to incorporate recently developed
MC radiative transfer techniques which allow the formation of such 
lines to be modelled.
In particular, {\it Macro Atoms}, as devised and
tested by Lucy (2002,2003), are used. 
This approach allows the radiative equilibrium
constraint to be rigorously enforced at all times 
without approximation in the treatment of interactions
between radiation and matter.

Initially, several MC simulations are performed to compute the temperature
and degrees of excitation and ionization throughout the wind. For these
``ionization cycles'' the MC quanta are launched with frequencies
spanning a wide enough range to simulate all the important radiative energy inputs
to the system. During each of these MC simulations, estimators for the 
radiative heating rate (e.g. due to photoionization) are recorded in each grid cell.
At the end of the simulation, these rates are balanced against cooling rates to determine
the local electron temperature which is then adopted in the next simulation. This 
process is repeated until essentially all grid cells pass 
a chosen convergence threshold in electron temperature.

For this work, the code has been modified to also record high
precision MC estimators for the individual radiative rates (photoionization
and bound-bound excitation) following Lucy (2003).
At the end of each ionization cycle, these
are used, together with the various collision rates and radiative decay 
rates, to compute level populations and the ionization fraction in each grid cell.
These populations are used in the next iteration, 
alleviating the need to use the approximate analytic formulae for ionization 
and excitation employed by Long \& Knigge (2002).
No convergence criterion is enforced for the level populations at present. Lucy (2002)
has shown that, provided the populations of the lowest levels are estimated sensibly, 
the indivisible-energy-packet MC method produces highly accurate emissivities even 
if the populations for the emitting upper levels are relatively poorly known.
This insensitivity motivates our use of the MC method since it substantially reduces
our reliance on detailed calculation of excited level populations for an 
adequate non-LTE investigation of recombination line formation.
We note that our convergence criterion on the electron temperature implies a corresponding
convergence in the energy flow between the radiation field and the thermal energy pool.
Since this energy flow is a summation of all the radiative heating processes in the hydrogen model,
its behaviour implies good convergence amongst all the processes which are energetically 
important.

After the ``ionization cycles'', a set
of additional MC simulations are performed to compute the spectrum
(the ``spectral cycles''). 
During these cycles the MC quanta are created 
in only a fixed frequency interval while
the thermal, ionization and excitation states of each grid cell are fixed to 
the values previously determined as described above. By tracking the quanta,
we compute the spectrum
as seen by observers at different inclination angles in the
frequency range of interest.

\begin{figure*}
{\large
\hspace{0.1cm}
$\theta = 30$\textdegree \hspace{1.64cm}
$\theta = 45$\textdegree \hspace{1.64cm}
$\theta = 60$\textdegree \hspace{1.64cm}
$\theta = 70$\textdegree \hspace{1.64cm}
$\theta = 80$\textdegree \hspace{1.64cm}
$\theta = 85$\textdegree \hspace{1.64cm}
}
\epsfig{file=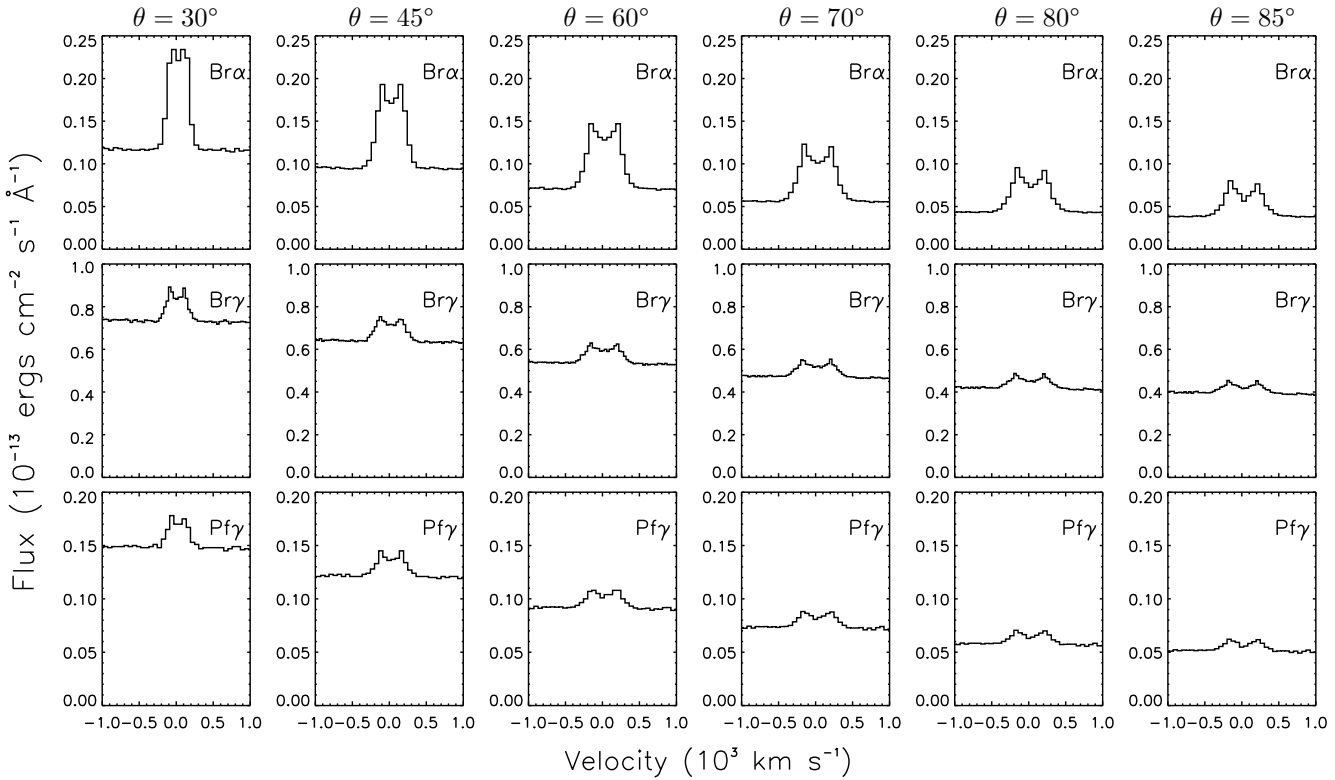, width=17cm}
\vspace{-2cm}
\caption{Model A (reference model with reflecting disk): computed Br$\alpha$ (upper), Br$\gamma$ (centre) and
Pf$\gamma$ (lower) line profiles for viewing angles
to the polar axis (from left to right) of 30\textdegree, 
45\textdegree, 60\textdegree, 70\textdegree, 
80\textdegree~ and 85\textdegree.
The flux is given for an unreddened source at a distance of 1~kpc 
and the velocity is measured relative to line centre.}
\end{figure*}

\begin{table}
\caption{The observed range of properties of massive YSO H~{\sc i} IR lines in 
the sample discussed by Bunn et al. (1995). Note that in several of the objects,
Br$\gamma$ and/or Pf$\gamma$ are not observed and so the range of EW
for these lines is biased towards the brighter sources compared with
Br$\alpha$.}
\begin{tabular}{lccc} \hline
Line & $-$EW & FWHM & HWZI \\
& (\AA) & km~s$^{-1}$ & km~s$^{-1}$\\ \hline
Br$\alpha$ &  4.8 -- 87 & 55 -- 155 & 50 -- 415\\
Br$\gamma$ &  2.4 -- 23 & 70 -- 150 & 70 -- 270\\
Pf$\gamma$ &  2.3 -- 12 & 110 -- 150 & 40 -- 260\\ \hline
& EWR  & &EWR \\
Br$\alpha$/Br$\gamma$ & 1.1 -- 4.4 & Br$\alpha$/Pf$\gamma$ & 3.0 -- 7.2\\ \hline
\end{tabular}
\end{table}

\begin{table*}
\caption{Equivalent widths (EW) and full-width-at-half-maximum (FWHM) for
the Br$\alpha$, Br$\gamma$ and Pf$\gamma$ lines at different viewing
angles ($\theta_{\mbox{\scriptsize obs}}$).
The values of FWHM have accuracy of around $\sim 50$~km~s$^{-1}$, limited 
by the frequency gridding of the calculations. The accuracy of the EWs is
limited by MC noise in the spectrum. Typically, the Monte
Carlo noise in the continuum is $< 2$ per cent.
}
\begin{tabular}{lcccccccccccc}\\ \hline
& \multicolumn{2}{c}{$\theta_{\mbox{\scriptsize obs}} = 30$\textdegree}
& \multicolumn{2}{c}{$\theta_{\mbox{\scriptsize obs}} = 45$\textdegree}
& \multicolumn{2}{c}{$\theta_{\mbox{\scriptsize obs}} = 60$\textdegree}
& \multicolumn{2}{c}{$\theta_{\mbox{\scriptsize obs}} = 70$\textdegree}
& \multicolumn{2}{c}{$\theta_{\mbox{\scriptsize obs}} = 80$\textdegree}
& \multicolumn{2}{c}{$\theta_{\mbox{\scriptsize obs}} = 85$\textdegree}\\
& $-$EW & FWHM & $-$EW & FWHM & $-$EW & FWHM & $-$EW & FWHM & $-$EW & FWHM & $-$EW & FWHM \\ 
Line & (\AA) & (km~s$^{-1}$) & (\AA) & (km~s$^{-1}$)  & (\AA) & (km~s$^{-1}$) 
& (\AA) & (km~s$^{-1}$) & (\AA) & (km~s$^{-1}$) 
& (\AA) & (km~s$^{-1}$) \\ \hline
\multicolumn{13}{l}{Model A (reference model with reflecting disk)}\\
Br$\alpha$ &  44 & 300 &  56 & 425 &  70 & 550 &
  78 & 610 &  77 & 610 &  68 & 550\\
Br$\gamma$ &  4.0 & 290 &  4.9 & 420 &  5.6 & 520 &
  5.8 & 520 &  5.8 & 550 &  5.4 & 550\\
Pf$\gamma$ &  7.2 & 340 &  9.3 & 450 &  12 & 560 &
  13 & 560 &  14 & 560 &  13 & 560 \\ \hline
\multicolumn{13}{l}{Model B (reference model with reprocessing disk)}\\
Br$\alpha$ &  3.1 & 430 &  4.2 & 610 &  6.1 & 730 &
  8.5 & 790 &  14 & 790 &  19 & 790\\
Br$\gamma$$^a$ &  0.17 & -- &  0.36 & -- &  0.41 & -- &
  0.57 & -- &  0.83 & -- &  1.2 & --\\
Pf$\gamma$$^a$ &  0.76 & -- &  0.80 & -- &  1.3 & -- &
  1.7 & -- &  2.8 & -- &  3.8 & -- \\ \hline
\multicolumn{13}{l}{Model C (as Model B but with mass loss rate enhancement)}\\
Br$\alpha$ &  27 & 300 &  38 & 425 &  56 & 550 &
  74 & 550 &  100 & 670 &  120 & 670\\
Br$\gamma$ &  3.4 & 290 &  4.8 & 420 &  7.1 & 520 &
  9.3 & 550 &  11 & 580 &  12 & 620\\
Pf$\gamma$ &  6.6 & 340 &  9.6 & 450 &  13 & 560 &
  17 & 560 &  21 & 560 &  22 & 670 \\ \hline
\multicolumn{13}{l}{Model D (as Model A but with disk out to 100~$r_{*}$)}\\
Br$\alpha$ &  24 & 300 &  32 & 425 &  43 & 550 &
  53 & 550 &  62 & 610 &  63 & 550\\
Br$\gamma$ &  3.4 & 360 &  3.9 & 420 &  4.5 & 550 &
  4.7 & 590 &  4.6 & 590 &  4.2 & 590\\
Pf$\gamma$ &  4.5 & 340 &  5.7 & 450 &  7.6 & 500 &
  9.5 & 560 &  11 & 560 &  12 & 560 \\ \hline
\multicolumn{13}{l}{Model E (as Model C but with disk out to 100~$r_{*}$)}\\
Br$\alpha$ &  4.6 & 300 &  6.6 & 430 &  10 & 550 &
  16 & 610 &  29 & 670 &  49 & 670\\
Br$\gamma$ &  0.80 & 390 &  1.1 & 550 &  1.6 & 650 &
  2.3 & 750 &  3.8 & 720 &  5.4 & 720\\
Pf$\gamma$ &  1.2 & 340 &  1.6 & 620 &  2.3 & 730 &
  3.1 & 730 &  5.1 & 730 &  8.1 & 730 \\ \hline
\multicolumn{13}{l}{Model F (disk with inner hole, see Section 4.4)}\\
Br$\alpha$ &  22 & 300 &  28 & 300 &  36 & 425 &
  43 & 425 &  51 & 425 &  52 & 490\\
Br$\gamma$ &  2.2 & 260 &  2.4 & 360 &  2.6 & 460 &
  2.7 & 490 &  2.7 & 490 &  2.5 & 490\\
Pf$\gamma$ &  2.8 & 220 &  3.8 & 340 &  5.0 & 450 &
  6.2 & 450 &  7.5 & 450 &  7.9 & 450 \\ \hline
\end{tabular}

\noindent $^a$ For Model~B, the Br$\gamma$ and Pf$\gamma$ lines are very 
weak and only appear marginally above the MC noise -- therefore, the
entries for these lines give only  
upper limits on their EW.

\end{table*}

\begin{table*}
\caption{Computed Br$\alpha$ line fluxes and flux ratios Br$\alpha$/Br$\gamma$ and
Br$\alpha$/Pf$\gamma$ at different viewing angles. Fluxes are given 
in units $10^{-13}$~ergs~cm$^{-2}$~s$^{-1}$
for an 
unreddened source at a distance of 1~kpc.}
\begin{tabular}{lcccccc}\\ \hline
&{$\theta_{\mbox{\scriptsize obs}} = 30$\textdegree}
&{$\theta_{\mbox{\scriptsize obs}} = 45$\textdegree}
&{$\theta_{\mbox{\scriptsize obs}} = 60$\textdegree}
&{$\theta_{\mbox{\scriptsize obs}} = 70$\textdegree}
&{$\theta_{\mbox{\scriptsize obs}} = 80$\textdegree}
&{$\theta_{\mbox{\scriptsize obs}} = 85$\textdegree}\\ \hline
\multicolumn{7}{l}{Model A (reference model with reflecting disk)}\\
Br$\alpha$ flux & 5.1 & 5.3 & 4.9 & 4.3 & 3.3 & 2.6\\
Br$\alpha$/Br$\gamma$ & 1.72 & 1.70 & 1.66 & 1.59 & 1.48 & 1.36 \\
Br$\alpha$/Pf$\gamma$ & 4.86 & 4.74 & 4.67 & 4.48 & 4.13 & 3.81 \\ \hline
\multicolumn{7}{l}{Model B (reference model with reprocessing disk)}\\
Br$\alpha$ flux & 4.8 & 5.0 & 4.5 & 3.9 & 3.1 & 2.6 \\
Br$\alpha$/Br$\gamma$$^a$ & 1.99 & 1.37 & 1.73 & 1.72 & 1.85 & 1.72\\
Br$\alpha$/Pf$\gamma$$^a$ & 3.07 & 4.09 & 3.34 & 3.84 & 3.65 & 3.77\\\hline
\multicolumn{7}{l}{Model C (as Model B but with mass-loss rate enhancement)}\\
Br$\alpha$ flux & 43 & 47 & 43 & 37 & 27 & 20 \\
Br$\alpha$/Br$\gamma$ & 0.91 & 0.93 & 0.93 & 0.95 & 1.05 & 1.19\\
Br$\alpha$/Pf$\gamma$ & 3.05 & 3.01 & 3.16 & 3.27 & 3.69 & 4.09\\ \hline
\multicolumn{7}{l}{Model D (as Model A but with disk out to 100~$r_{*}$)}\\
Br$\alpha$ flux & 4.0 & 4.1 & 4.0 & 3.6 & 3.0 & 2.5 \\
Br$\alpha$/Br$\gamma$ & 1.54 & 1.59 & 1.60 & 1.59 & 1.56 & 1.52\\
Br$\alpha$/Pf$\gamma$ & 4.50 & 4.61 & 4.66 & 4.49 & 4.33 & 4.18\\ \hline
\multicolumn{7}{l}{Model E (as Model C but with disk out to 100~$r_{*}$)}\\
Br$\alpha$ flux & 25 & 26 & 25 & 23 & 19 & 16\\
Br$\alpha$/Br$\gamma$ & 1.32 & 1.40 & 1.49 & 1.51 & 1.56 & 1.64\\
Br$\alpha$/Pf$\gamma$ & 3.07 & 3.44 & 3.66 & 4.11 & 4.57 & 4.90\\ \hline
\multicolumn{7}{l}{Model F (disk with inner hole, see Section 4.4)}\\
Br$\alpha$ flux & 19 & 19 & 19 & 17 & 16 & 14\\
Br$\alpha$/Br$\gamma$ & 2.37 & 2.37 & 2.29 & 2.18 & 2.06 & 2.03\\
Br$\alpha$/Pf$\gamma$ & 6.94 & 6.20 & 5.82 & 5.60 & 5.17 & 4.93\\ \hline
\end{tabular}

\noindent $^a$ For Model~B, the Br$\gamma$ and Pf$\gamma$ lines are very 
weak and only appear marginally above the MC noise - therefore, the
entries for these lines give only  
lower limits on the appropriate flux ratios.
\end{table*}

In all the calculations described below, the model structure and
radiation field properties are discretized onto a two-dimension ($r$-$z$) grid consisting of 
3000 cells. During each MC calculation, the input radiative energy (which arises
from the star and accretion disk) is divided into $3 \times 10^6$ indivisible quanta.

We have explicitly verified that our approach is adequate for the calculation of the spectrum
by comparing results for models with identical input parameters 
using smaller and larger numbers of grid cells. 
These calculations differed by at most a few per cent in the
computed H~{\sc i} line strengths from the standard (3000 cell) case. This implies that the 
number of packets, number of grid cells and the quality of the MC estimators 
are all sufficient for calculations to this degree of accuracy.

\section{Results}

\subsection{Observational constraints}

Early observations of IR H~{\sc i} emission lines in massive YSOs (e.g. Simon et al. 1981)
had insufficient signal-to-noise ratio to provide detailed information on the line shapes, thus
subsequent theoretical work (e.g. H\"{o}flich \& Wehrse 1987) was concerned primarily
with modelling the observed line fluxes and flux ratios. 

More recently, higher quality data (e.g. Drew et al. 1993, Bunn et al. 1995) has allowed
the line shapes to be studied in greater detail in several massive YSOs.
In common with earlier work (e.g. Simon et al. 1981,1983; Drew et al. 1993),
Bunn et al. (1995) observed a moderately strong, pure-emission line of Br$\alpha$ in all of their
targets. In most cases, they also reported weaker emission in Br$\gamma$ and Pf$\gamma$.
Their observed line profiles are single peaked and not shifted from the expected local rest velocity
by more than a few 
km~s$^{-1}$. Many of the profiles show line wings which extend to several hundred km~s$^{-1}$
from line centre and in some cases these wings are clearly asymmetric. By considering 
wavelength-dependent line ratios (following Drew et al. 1993), Bunn et al. (1995) 
found evidence, in at least some cases, of hybrid profiles consisting of a narrow line core
(which may have a nebular origin) and broad wings in which the line ratio is consistent with
formation in an accelerating outflow.

High resolution observations of several massive YSOs have very recently been presented by 
Blum et al. (2004). Interestingly, in their sample, at least one object (NGC~3576) 
has double-peaked,
narrow line profiles. However, it is difficult to study the broad line wings in the Blum et al. (2004) 
data since the spectral coverage is too narrow to establish the underlying continuum with certainty.

\begin{figure}
\epsfig{file=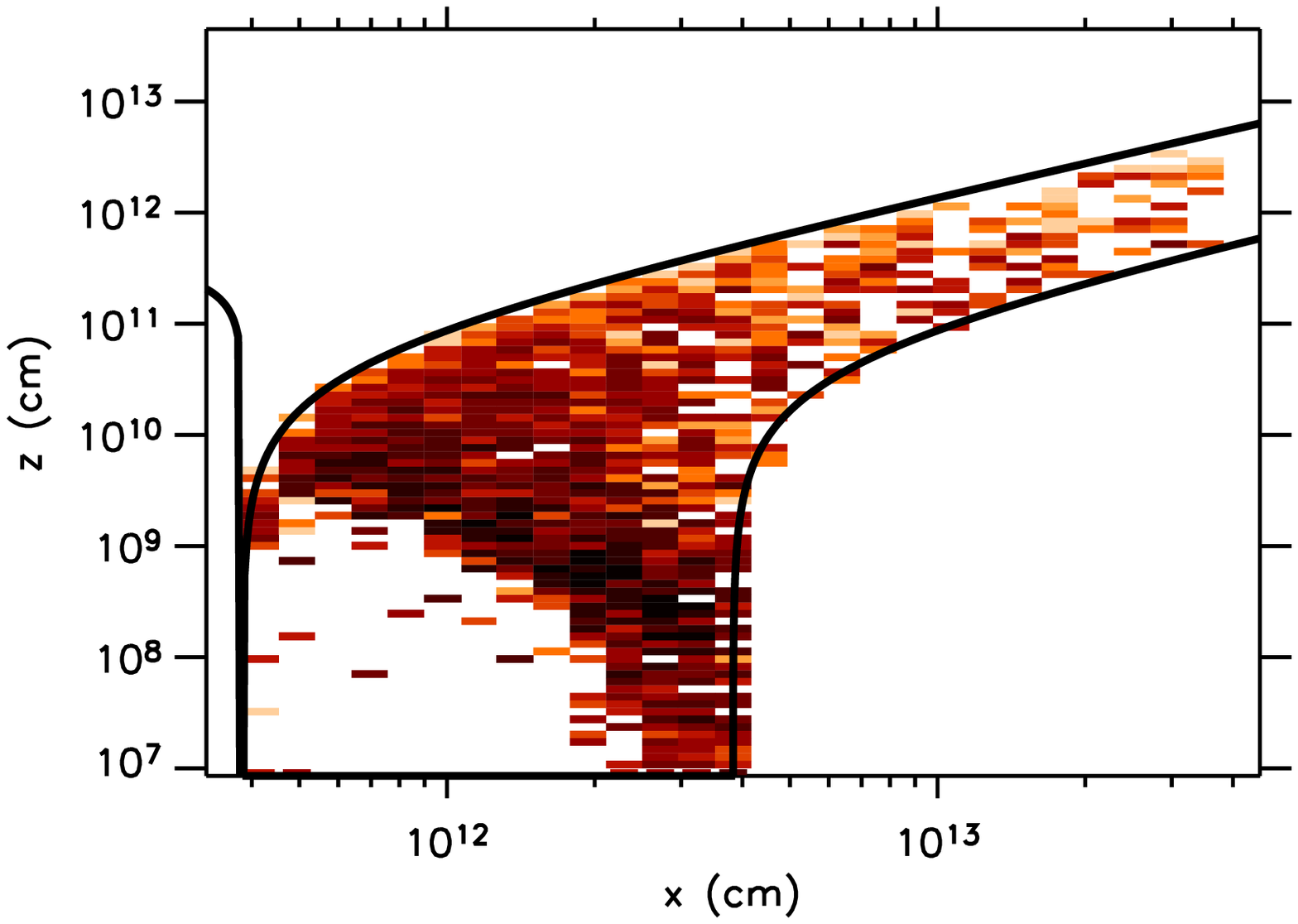, height=5.5cm}
\epsfig{file=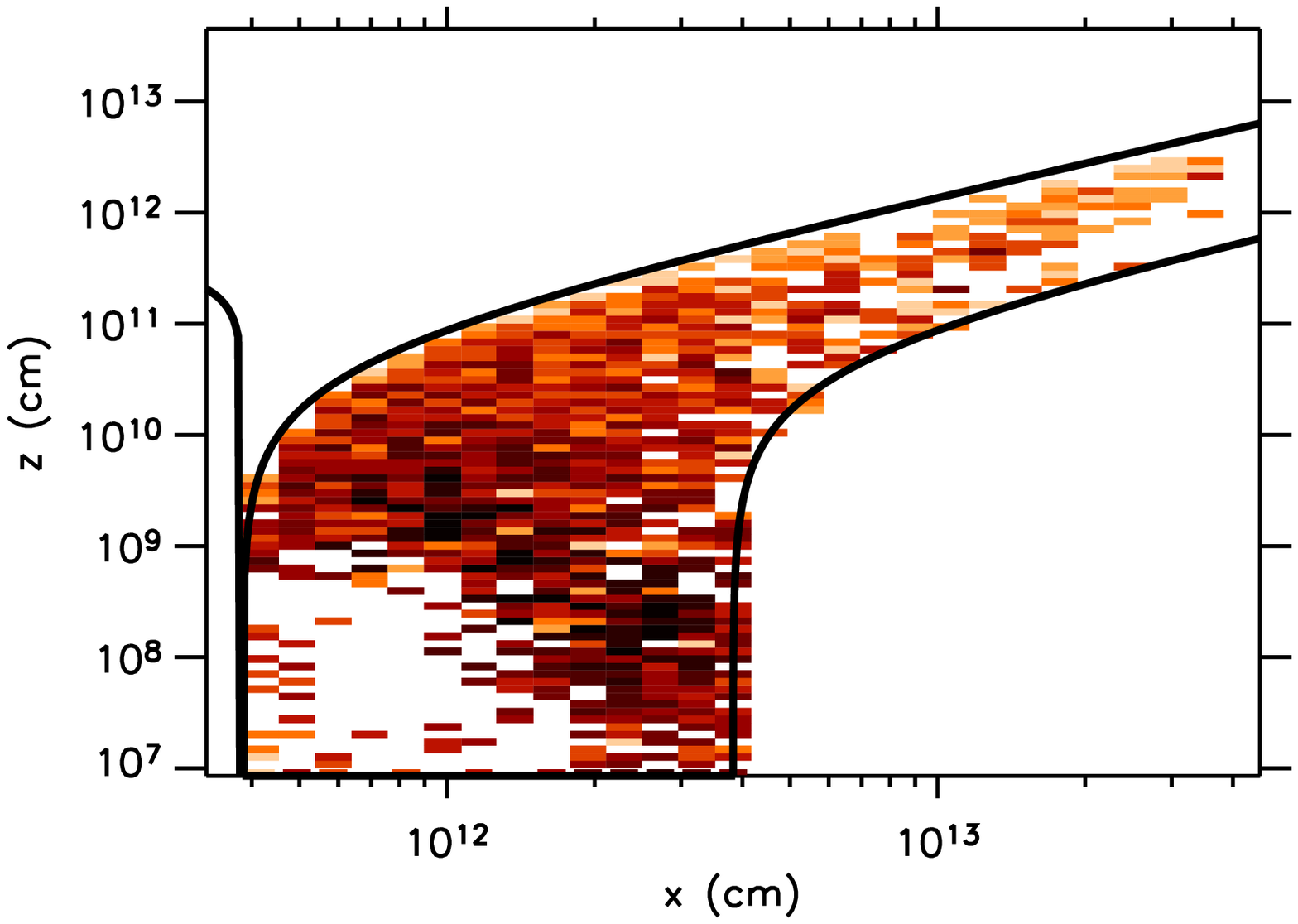, height=5.5cm}
\epsfig{file=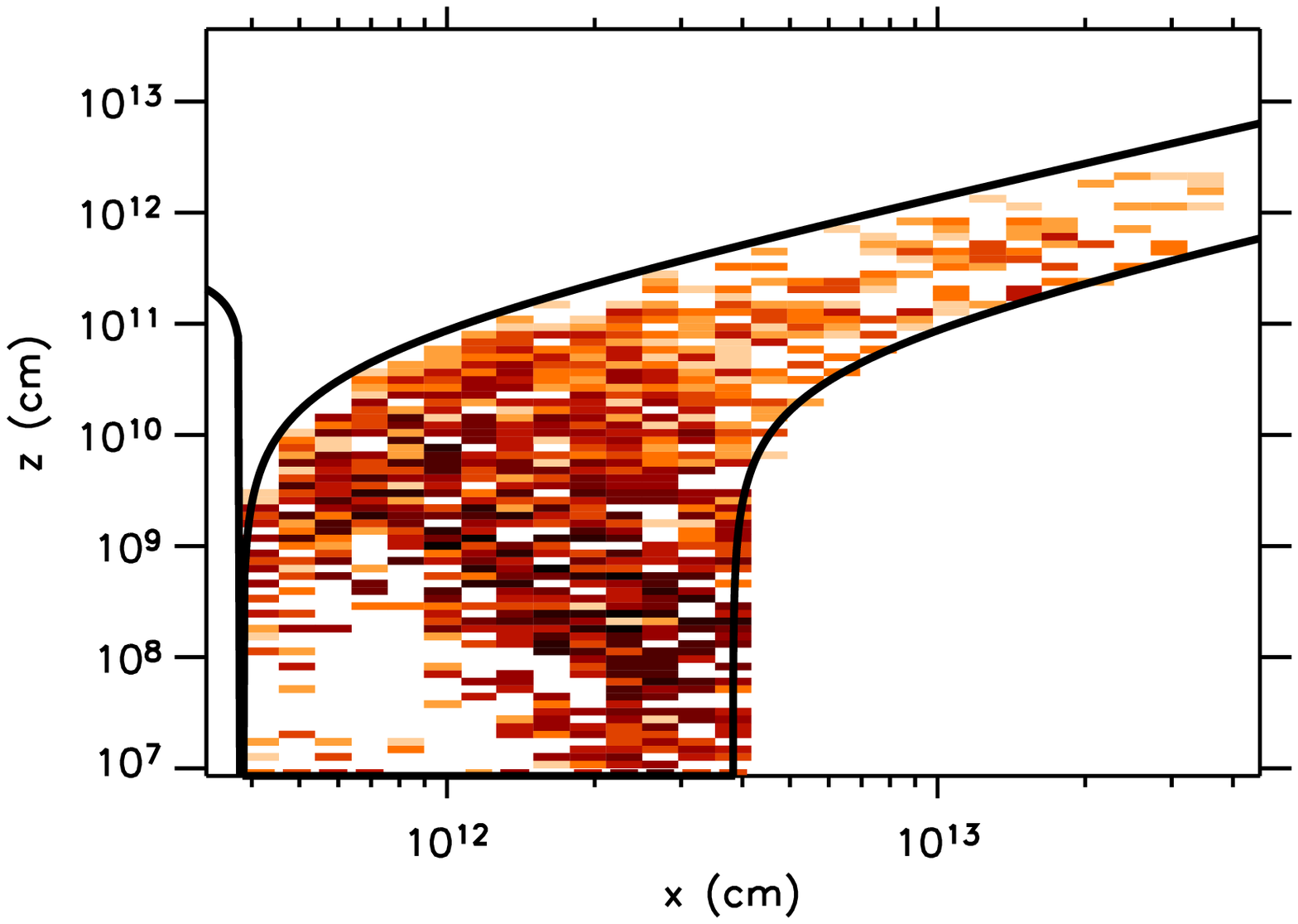, height=5.5cm}
\caption{Regions of line formation for 
Br$\alpha$ (upper), Br$\gamma$ (centre) and
Pf$\gamma$ (lower). Note the logarithmic axes.
The solid black lines define the boundaries of the disk wind (the star surface, the disk surface
and the inner and outer wind boundaries defined by 
$\theta_{\mbox{\scriptsize min}}$ and
$\theta_{\mbox{\scriptsize max}}$). Within the wind, the volume is divided into 
the computational grid cells which are shaded proportional to the energy escaping to infinity
from that cell in the spectral line under consideration. The shading scale is logarithmic. The graininess
is due to MC noise.}
\end{figure}

In the following sub-sections, results of our models will be compared primarily with the
observations presented by Bunn et al. (1995) since their spectra provide good constraints on both
line strengths and shapes for a significant sample of objects.
Unfortunately, owing to the great
difficulty in establishing the extinction along the line of sight to massive YSOs, interpreting 
line fluxes and flux ratios for lines at different wavelengths is very difficult (it is known 
from Br$\alpha$/Pf$\gamma$ flux ratios that the H~{\sc i} lines are not well described by 
either the LTE or the Baker Menzel case B recombination assumptions -- thus the Br$\alpha$/Br$\gamma$
ratio cannot be trusted as an indicator of the reddening [H\"{o}flich \& Wehrse 1987]).
Therefore, in our discussion, we prefer to focus on EWs, EW ratios (EWRs) and line shapes 
since these are not affected by reddening.
For quantitative comparison, Table~2 gives a summary of the range of
observed EW, FWHM and half-width-at-zero-intensity (HWZI) from Bunn et al. (1995).
The table also gives the
observed range of EWRs for Br$\alpha$/Br$\gamma$ and
Br$\alpha$/Pf$\gamma$\footnote{Hereafter, these EWRs will be referred to as
EWR[Br$\alpha$/Br$\gamma$] and EWR[Br$\alpha$/Pf$\gamma$] respectively.}.
Since our model 
parameters are not fine-tuned to one particular object, 
these values should be regarded only as indicative of
the appropriate regime in which acceptable model predictions should lie.

\subsection{Reference model (Model A)}

Spectra have been computed in the 2 -- 5 $\mu$m range using the reference
wind model (Model A) which was described in 
Section~2. In this 
wavelength range, the model predicts moderately
strong emission in the first three lines of the Brackett series and 
weaker emission in lines of the Pfund series (Pf$\beta$ and Pf$\gamma$).
This is in qualitative agreement with the observed IR properties of 
massive YSOs (e.g. Simon et al. 1981,1983; Drew et al. 1993; Bunn et al. 1995).

Figure~5 shows the predicted Model~A line profiles of 
Br$\alpha$, Br$\gamma$ and Pf$\gamma$. Spectra are shown for a grid of
viewing angles ranging from 30\textdegree~to 85\textdegree~(measured 
relative to the polar axis).
For quantitative comparison with observations,
the computed EW and FWHM of these lines are tabulated
in Table~3. For completeness, computed line fluxes and flux ratios are given in Table~4.

For all three lines, the EW is largest for viewing angles ($\theta_{\mbox{\scriptsize obs}}$) of
70 -- 80\textdegree~and becomes noticeably smaller at lower 
$\theta_{\mbox{\scriptsize obs}}$.
The computed Br$\alpha$ EWs are consistent with 
the high end of the
range of measured EW (Table~2) obtained by Bunn et al. (1995). Similar agreement
between observations and the model
is also found for Pf$\gamma$ making the modelled 
EWR[Br$\alpha$/Pf$\gamma$] similar to the observed ratio.

For Br$\gamma$, the Model~A computed EWs lie in the mid-range
of the observed values (see Table 2). 
The relative weakness of the computed Br$\gamma$ line means that the
model EWR[Br$\alpha$/Br$\gamma$] is systematically too high --
in the observations the EWR for these lines ranges 
from 1.1 to 4.4 but the computed Model~A ratio is somewhat higher, ranging
from 11 to 13 for the viewing angles considered.
It is probable that part of this 
disagreement is the result of opacities being too 
low in the models -- but given the good agreement with observations for  
EWR[Br$\alpha$/Pf$\gamma$], it seems unlikely that a simple 
underestimate of the opacity in Br$\alpha$ is solely responsible. 
Perhaps more importantly,
the EWR[Br$\alpha$/Br$\gamma$] will be affected by 
inaccuracies in the model
continuum shape -- since Br$\alpha$ and Pf$\gamma$ lie at similar 
wavelengths, the choice of continuum shape will have 
little effect  on the computed ratio of their EWs but because
Br$\gamma$ lies at a significantly shorter wavelength, the 
EWR[Br$\alpha$/Br$\gamma$] 
is much more sensitive.
An alternative treatment of the continuum -- that invoking a
reprocessing disk -- is considered in Section~4.2.
It is also noted that, compared with Br$\alpha$ and Pf$\gamma$, 
observations of Br$\gamma$ for heavily 
embedded targets are more likely to be contaminated by emission 
from other sources
(e.g. larger scale nebular emission) owing to the wavelength dependence of the source
extinction.

The computed line profiles are complex, usually exhibiting double-peaks which become
more pronounced at larger $\theta_{\mbox{\scriptsize obs}}$.
This is in contrast to the observations where only single-peaked 
profiles are usually observed (see Section 4.1). Also, the computed profiles are too broad (FWHM
ranging from 300 to 600 km~s$^{-1}$ as a function of 
$\theta_{\mbox{\scriptsize obs}}$). 
Part of this discrepancy between
model and observations may be
explained by contamination of the observed profiles by narrow
nebular emission -- there is some evidence for hybrid line profiles with narrow cores  (see Section 4.1) -- 
however, the 
half-width-zero-intensity (HWZI) measurements also suggest that the observed line
profiles do not extend more than 200 to 300 km~s$^{-1}$ from line-centre
in most cases.

The large line widths and double-peaked profiles arise due to the rotational velocities in the 
region of line formation. Figure~6 identifies the region of line formation for the
Br$\alpha$, Br$\gamma$ and Pf$\gamma$ lines in Model~A; specifically, the
plots show the amount of energy that escapes in each spectral line
from each grid cell in the model. Note that MC noise is responsible for the graininess
of the figures. It can be seen that in all three cases, the region of line formation 
is close to the disk surface, within the first few velocity law scale-lengths 
along the wind streamlines. Br$\alpha$ tends to form slightly further out than either Br$\gamma$
or Pf$\gamma$ -- this is to be expected owing to the higher opacity in the Br$\alpha$ line.

To obtain models which produce narrower lines, either more low-velocity gas must be added, 
or high-velocity gas must be removed. The simplest way in which to introduce low-velocity gas is by
increasing the outer disk radius ($r_{\mbox{\scriptsize disk}}$) -- models with 
wider disks are presented in Section 4.3. To remove high-velocity gas 
requires a reduction in the wind density at small values of $r$. Given that the inner disk
is brighter than the outer disk, it seems unlikely that a radiatively driven flow will not have greatest
mass-loading at small radii. Therefore, the most plausible way in which to eliminate high-velocity gas
is not to change the prescription of mass-loading but rather to consider models in which either the
stellar radius is larger than that adopted in the reference model or the disk
inner radius is several times greater than the adopted stellar radius.
An example of the latter,
a model
with an empty inner cavity in the disk, is discussed in Section~4.5.

To assess the relative importance of the disk wind and stellar wind component,
results from a calculation considering only the disk wind component were compared with
those which include the stellar wind. These two calculations give virtually
identical results -- this is not unexpected since it is already known that
a spherical wind would require a mass-loss rate at least an order of magnitude greater 
than that in the stellar wind component here (Simon et al. 1981, 
see Section 1). In view of this, in all the calculations discussed below, the
stellar wind component is dropped for computational efficiency.

\subsection{Reprocessing of radiation by the disk (Models B and C)}

In Model~A (discussed above) it was assumed that the photons striking the
accretion disk were reflected rather than absorbed. 
To investigate an alternative simple hypothesis, namely that the disk
absorbs and thermally re-emits
the radiation which strikes it 
(see Section 2.2.2), a second model (Model B) is now considered.
This is the extreme, optically thick disk case.
All the system parameters (Table~1) for Model~B are identical to those of 
Model~A, saving that the stellar wind component is omitted (see above).

The simple treatment of reprocessing adopted (see Section 2.2.2) means that
the disk is much brighter at IR wavelengths in Model B than Model A. Thus,
although the wind parameters in Model B are unchanged, the EWs
of all the emission lines are much smaller 
(the EWs are given in Table~3). At moderate to high
$\theta_{\mbox{\scriptsize obs}}$, the Model~B EWs for
Br$\alpha$ and Pf$\gamma$ are comparable to the observed EWs
in the weaker sources reported by Bunn et al. (1995). 
However, the computed 
Br$\gamma$ line is very weak.

To increase the line EWs in the presence of a reprocessing disk
the wind density (and hence emission measure) needs
to be increased. 
Therefore,
a further model (Model~C) has been constructed -- it is 
identical to Model B saving that the disk-wind mass-loss rate has been
increased by a factor of four. 
A higher mass-loss rate may be feasible 
even under the assumption of driving purely by radiation
since the
absolute mass-loss rates provided by the hydrodynamical model of Drew et 
al. (1998) may be uncertain by a factor of a few owing to the 
parameterisation of the line-driving force.
EWs and FWHM for Model C are
given in Table~3 and the line profiles computed from this model are shown in 
Figure~7.

The Model~C EWs are slightly greater than even the largest 
observed values from the Bunn et al. (1995) data, suggesting that the 
factor of four mass-loss rate increase between models B and C comfortably
brackets the range of values that are likely to be consistent with the data
for particular objects. 

The EWR[Br$\alpha$/Pf$\gamma$] 
is smaller for Model~C than Model~A,
lying close to the mid-range of observed values.
The EWR[Br$\alpha$/Br$\gamma$] is also smaller in Model~C -- it is still
greater than the observed ratio in any of the objects observed by Bunn et al. (1995)
but the discrepancy is much less.
This change in the line ratios is mostly due to the increased Br$\alpha$ opacity
resulting from the higher wind density. This result suggests that by further 
increasing the mass-loss rate an EWR[Br$\alpha$/Br$\gamma$] close to that
observed could be achieved. However, it is noted that a further increase
in the wind density will lead to Br$\alpha$ EWs which are
uncomfortably large.

The Model~C line profiles are generally less 
double-peaked and more square-topped than the Model~A profiles.
Their FWHM are very similar to those in Model~A and the line wings
still extend to several hundred km~s$^{-1}$ at the base.

In general, Model~C appears to suggest that the combination of a reprocessing 
disk and a higher disk-wind mass-loss rate is in closer agreement with the 
observations than the reference model (Model A). But further hydrodynamical modelling
is required to examine whether so high a mass-loss rate is feasible.

\subsection{The influence of the outer radius of the disk}

The outer disk radius paralleling that of Drew et al. (1998) and 
adopted in Models A -- C
is, arguably, rather small. 
In principle, line formation in a wind
rising from the outer parts of a larger disk could eliminate the double-peaked
line shape obtained with Models A -- C by adding low-velocity gas.
In this section two more models (Models D and E)
are considered: these models are identical, respectively, to Models A and C
saving that the outer disk radius is set to 100~$r_{*}$ rather than 
10~$r_{*}$.

The computed line profiles for Model D are shows in Figure~8 and the
EWs for Models D and E are given in Table~3. The differences
between the profile shapes shown in Figures 5 and 8 are generally fairly 
small. The double-peaked shape is slightly filled in when the larger 
disk is considered; however, the assumption that the mass-loading of streamlines is 
proportional to the luminous flux (see Section 2.2.3) means that the majority of 
the mass-loss still originates from the inner disk where velocities are high.
The most important consequence of the larger disk is the increase in the IR continuum
level (see Figure~3 and the discussion in Section~2.3.2). The brighter continuum
means that, when the disk is bigger, the line EWs are all smaller. Also,
the EWR[Br$\alpha$/Br$\gamma$] is smaller -- the 
larger disk radius increases the continuum around 4~$\mu$m by more than at 2~$\mu$m (see Figure 3).

For Model~D, the Br$\alpha$ EWs lie comfortably in the mid-range
of the observed values and the EWR[Br$\alpha$/Pf$\gamma$] is similar to
that found in Model~A. The EWR[Br$\alpha$/Br$\gamma$] is smaller in Model~D than
A (as required) but the improvement is less significant than obtained with Model~C (see 
Section 4.2).
Model~E predicts smaller EWs, but still within the range observed.
The EWR[Br$\alpha$/Br$\gamma$] in Model~E is closer to that observed than in any 
of the other models -- this follows since Model~E benefits from the improvement 
in this ratio resulting from both a higher line opacity (density) and stronger
disk continuum.

\subsection{The influence of a central hole in the disk}

\begin{figure*}
{\large
\hspace{0.1cm}
$\theta = 30$\textdegree \hspace{1.64cm}
$\theta = 45$\textdegree \hspace{1.64cm}
$\theta = 60$\textdegree \hspace{1.64cm}
$\theta = 70$\textdegree \hspace{1.64cm}
$\theta = 80$\textdegree \hspace{1.64cm}
$\theta = 85$\textdegree \hspace{1.64cm}
}
\epsfig{file=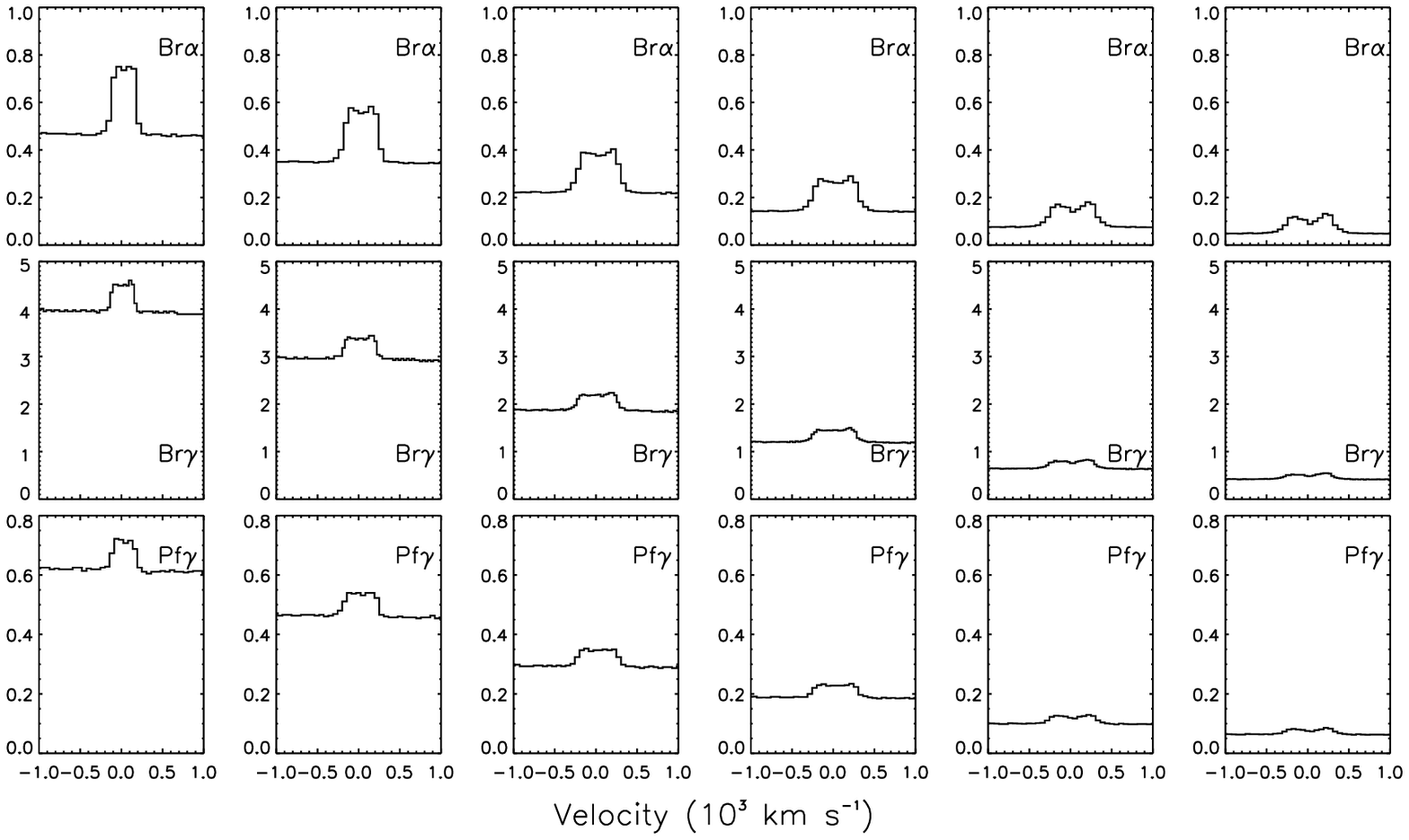, width=17cm}
\vspace{-2cm}
\caption{As Figure 5 but showing Model C (reprocessing disk, enhanced mass-loss rate) results.}
\end{figure*}

\begin{figure*}
{\large
\hspace{0.1cm}
$\theta = 30$\textdegree \hspace{1.64cm}
$\theta = 45$\textdegree \hspace{1.64cm}
$\theta = 60$\textdegree \hspace{1.64cm}
$\theta = 70$\textdegree \hspace{1.64cm}
$\theta = 80$\textdegree \hspace{1.64cm}
$\theta = 85$\textdegree \hspace{1.64cm}
}
\epsfig{file=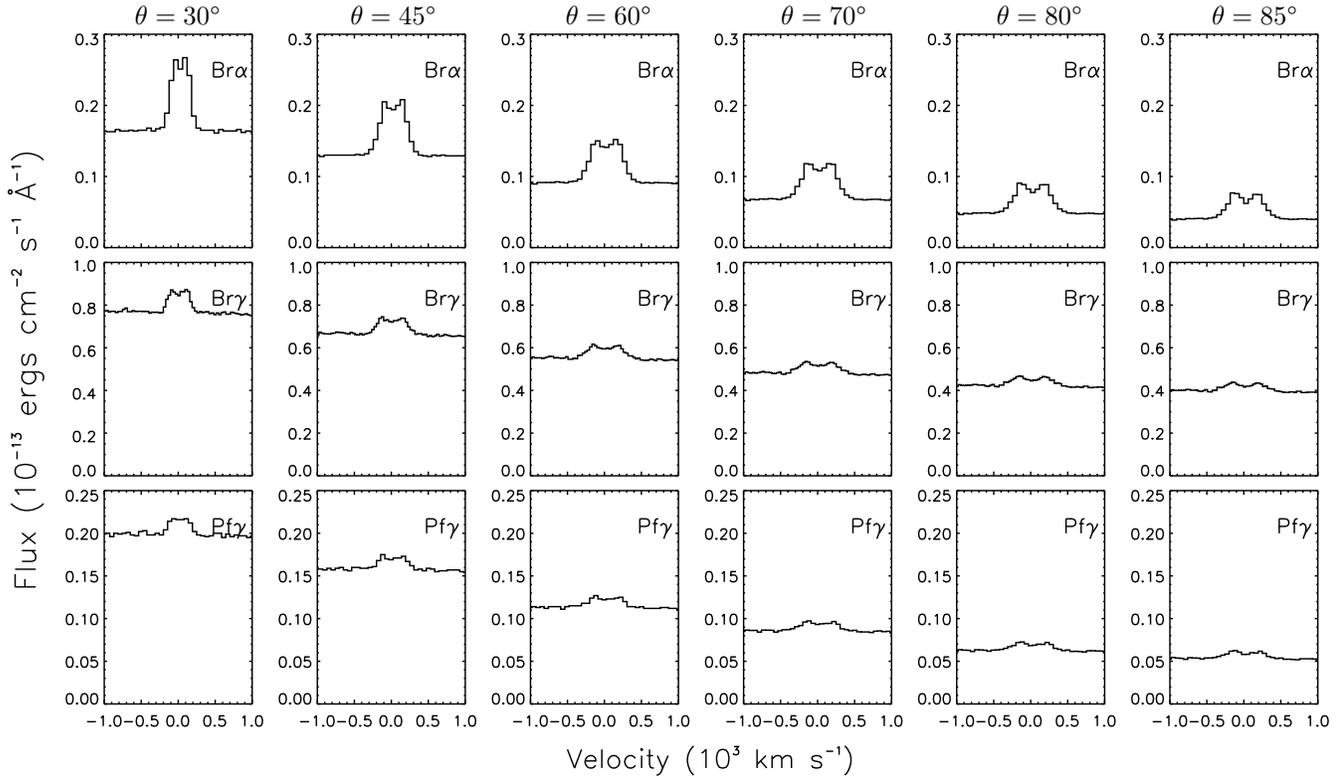, width=17cm}
\vspace{-2cm}
\caption{As Figure 5 but showing Model D (reflecting disk with larger outer radius) results.}
\end{figure*}

\begin{figure*}
{\large
\hspace{0.1cm}
$\theta = 30$\textdegree \hspace{1.64cm}
$\theta = 45$\textdegree \hspace{1.64cm}
$\theta = 60$\textdegree \hspace{1.64cm}
$\theta = 70$\textdegree \hspace{1.64cm}
$\theta = 80$\textdegree \hspace{1.64cm}
$\theta = 85$\textdegree \hspace{1.64cm}
}
\epsfig{file=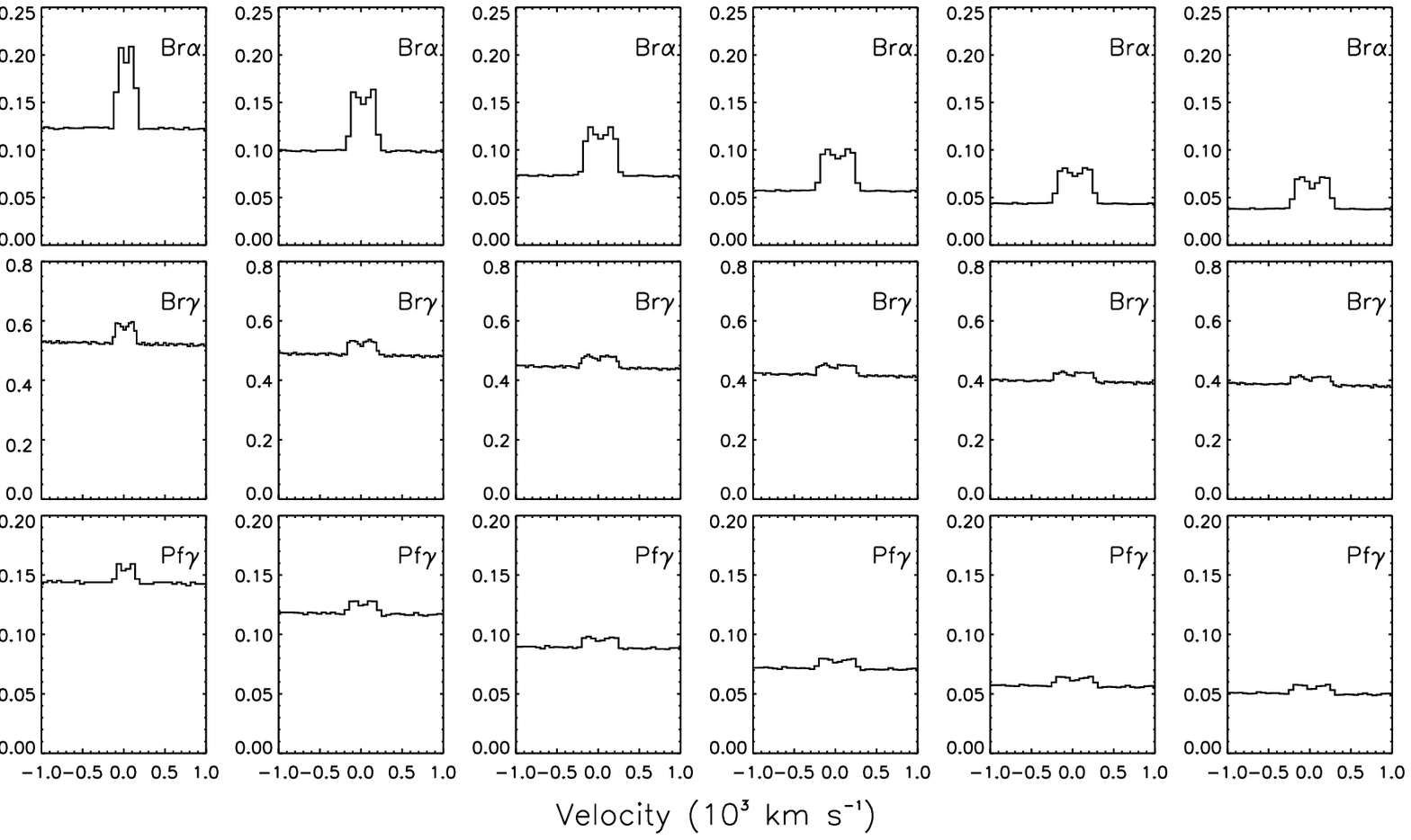, width=17cm}
\vspace{-2cm}
\caption{As Figure 5 but showing Model F (disk with inner hole) results.}
\end{figure*}

It is apparent that a failing of Models A -- E in comparison with 
the data is that the computed line profiles are too broad. 
One of two possible solutions to this 
problem is to invoke an inner hole in the accretion
disk, thereby removing the highest velocity material from the region of
line formation. Optically thin inner cavities in accretion disks around 
young luminous stars have previously been suggested in various contexts
(e.g. Herbig Ae/Be stars -- Dullemond et al. 2001; Tuthill, 
Monnier \& Danshi 2001; 
Natta et al. 2001; and see Monnier et al. 2005 for a recent discussion).
Interestingly, a disk with an inner hole might also help explain the 
residual discrepancy between the observed EWR[Br$\alpha$/Br$\gamma$] and
the models discussed above. Since the inner parts of the accretion disk are
hottest, removing them will push the SED of the 
disk to the red. This will increase the continuum level at Br$\alpha$ relative to
Br$\gamma$ and therefore reduce the EWR[Br$\alpha$/Br$\gamma$] as required.
However, a quantitative investigation of this possibility goes beyond the scope of this
paper since more sophisticated models for the disk emission than used here would be
required.

Consideration of models in which the disk has a central hole
represents a significant departure from the Drew et al. (1998) model and 
therefore, until further hydrodynamical calculations are performed, it
is difficult to place useful constraints on the likely geometry and parameters for
mass-loss from such a disk. 
Nevertheless, 
for illustrative purposes, a model adopting the same wind opening angles
($\theta_{\mbox{\scriptsize min}}$, $\theta_{\mbox{\scriptsize max}}$) and
mass-loss rate as the reference model (Model A; Section 4.1) but with an 
inner disk radius of $r_{\mbox{\scriptsize in}} = 5$~$r_{*}$ and outer
disk radius of $r_{\mbox{\scriptsize in}} = 100$~$r_{*}$ has been constructed
(Model~F) -- this corresponds to a disk with a circular inner hole of
$\sim 0.1$~AU. 
It is noted that, conceptually, invoking a wind launched significantly further out
than the stellar surface is reminiscent of the photoevaporation model (e.g. Hollenbach 
et al. 1994) -- however, the distance from the central star at which the 
mass-loss occurs here is still
much smaller than expected from photoevaporation. It is, however, comparable to the 
smallest disk radii inferred from CO observations 
by Bik \& Thi (2004).

The computed FWHM and EWs for Model~F are
tabulated in Table~3 and the profiles are shown in Figure~9.
The EWs are smaller than those for Model~A 
because the density in the wind is lower, a result of the radial 
dependence of the volume element in 
a cylindrical polar system. The lower density
also leads to lower opacity and hence larger ratios of Br$\alpha$ to Br$\gamma$ and 
Pf$\gamma$. However, as anticipated, the Model~F lines are narrower
than in Model~A, by up to 100~km~s$^{-1}$ in FWHM for moderate inclination 
angles. The profile widths at base are also smaller -- in Model~F the profiles
cut-off at $\sim 300$~km~s${^{-1}}$ from line centre. Thus holes of the size considered
here, or a little larger, could explain why the HWZI measurements of Bunn et al. (1995)
are rather smaller than would be expected based on Models~A -- E.

Given that there are no strong constraints on the hole size, geometry or 
mass-loss rate appropriate for the model here it is not possible to drawn
definitive conclusions.
However, it seems that models with an inner disk cavity
have the potential to resolve the discrepancies between the computed
disk wind line profiles and the observations. 
Model~F also suggests that the second possibility mentioned in Section~4.2 -- that
of an enlarged stellar radius -- can also provide a possible solution since increasing
the stellar radius will affect the line widths in a similar way to introducing an inner
gap in the disk.
Constraints on
the size of any inner hole, the stellar radius and the geometry of the disk wind 
are needed in order to proceed. 
These could be provided via
sophisticated observational 
techniques 
such as interferometry (which has already been applied
to Herbig Ae/Be stars by e.g. Monnier et al. 2005) or spectropolarimetry
(as discussed by Vink, Harries \& Drew 2005).

\section{Discussion}

As mentioned in Section~4,
direct confrontation of theoretical line fluxes with observations is difficult owing to 
the uncertainty in the extinction along lines-of-sight to massive YSOs. 
But on
comparing EWs -- the quantities most readily extracted from observations -- 
the radiative transfer calculations presented above lead us to conclude
that model disk winds, such as that 
obtained by Drew et al. (1998), do predict H~{\sc i} line strengths 
consistent with observations of massive YSOs.  This matching is achieved for total mass-loss
rates between 0.3 and 1.2 $\times 10^{-7}$ M$_{\odot}$~yr$^{-1}$,  in contrast with
previously calculated  
spherical models for early B stars which require mass loss rates up to $10^{-6}$ 
M$_{\odot}$~yr$^{-1}$ (Simon et al 1983, Nisini et al 1995).

There are, however, some difficulties when the models are considered in detail.
Although the computed EWR[Br$\alpha$/Pf$\gamma$] agrees well with observations,
the EWR[Br$\alpha$/Br$\gamma$] tends to be overpredicted by the models. 
We have investigated the effect of a higher density on this ratio
by considering models with greater mass-loss rate (e.g. Model~C; recall that
the model density is sensitive to several parameters in addition to the mass-loss 
rate and thus there is a degree of degeneracy among these quantities -- see Section
2.3.5).  As expected, increasing the wind density  decreases 
the EWR[Br$\alpha$/Br$\gamma$] owing to the greater optical depth. However, to 
fully explain the EWR discrepancy solely by such a modification is difficult since the 
absolute values of the
EWs grow rapidly if the density is raised.  At the same time, this EWR 
is very sensitive to the assumed underlying SED, owing to the significant wavelength
difference between the Br$\alpha$ and Br$\gamma$ lines (unlike Br$\alpha$ and Pf$\gamma$
which lie at similar wavelengths).   Observationally it is clear that
there is genuine diversity in the IR SEDs of massive YSOs (e.g. Henning et al. 1990).
In our models we have considered only two limiting cases for the treatment of the SED and
have not concerned ourselves with trying to fit specific objects.  To determine whether
the EWR[Br$\alpha$/Br$\gamma$ discrepancy is unavoidable, there is now a need to move
on to fitting observations of individual well-observed sources with reliable reddening
estimates.

\begin{figure*}
\epsfig{file=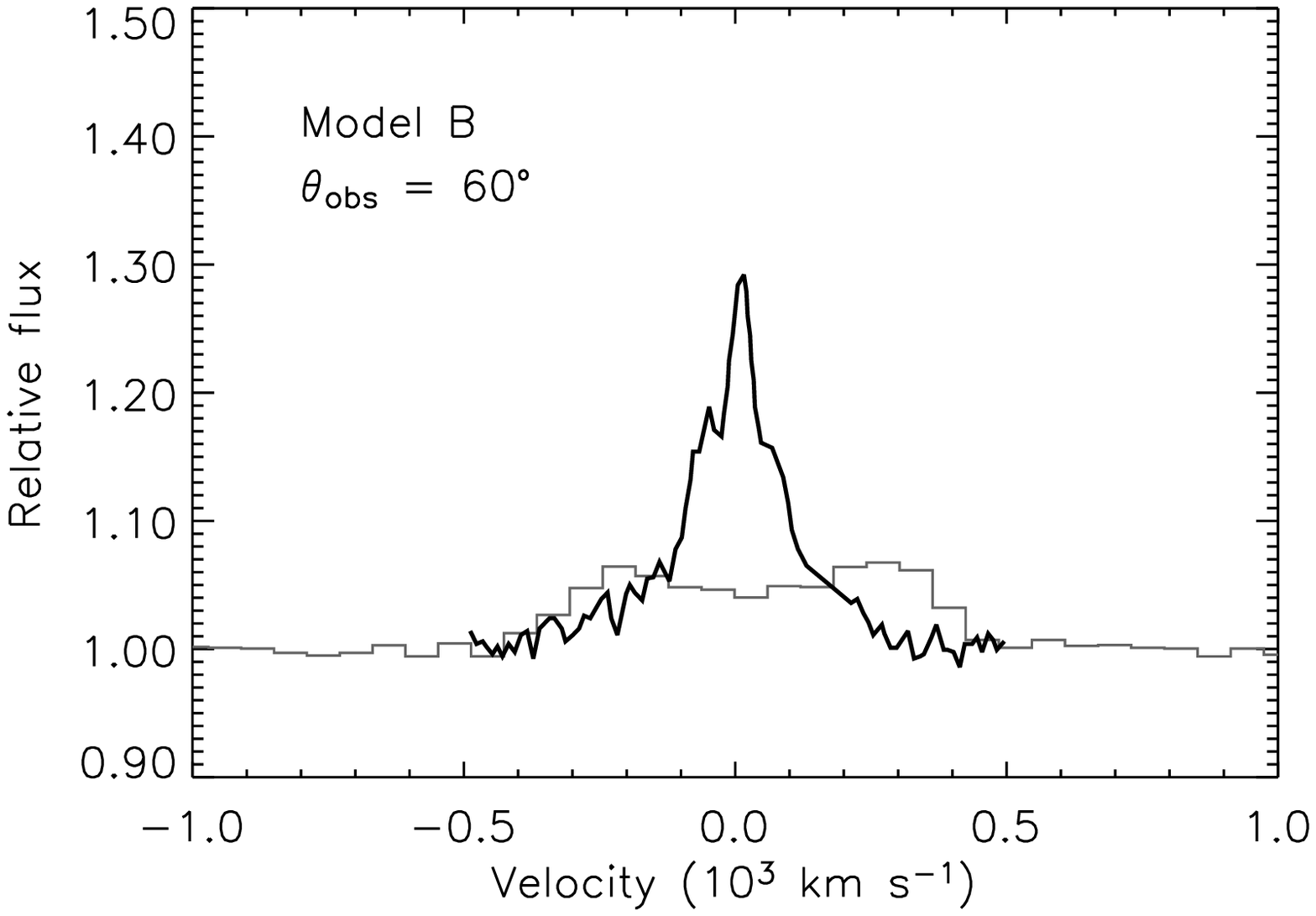, width=7cm}
\epsfig{file=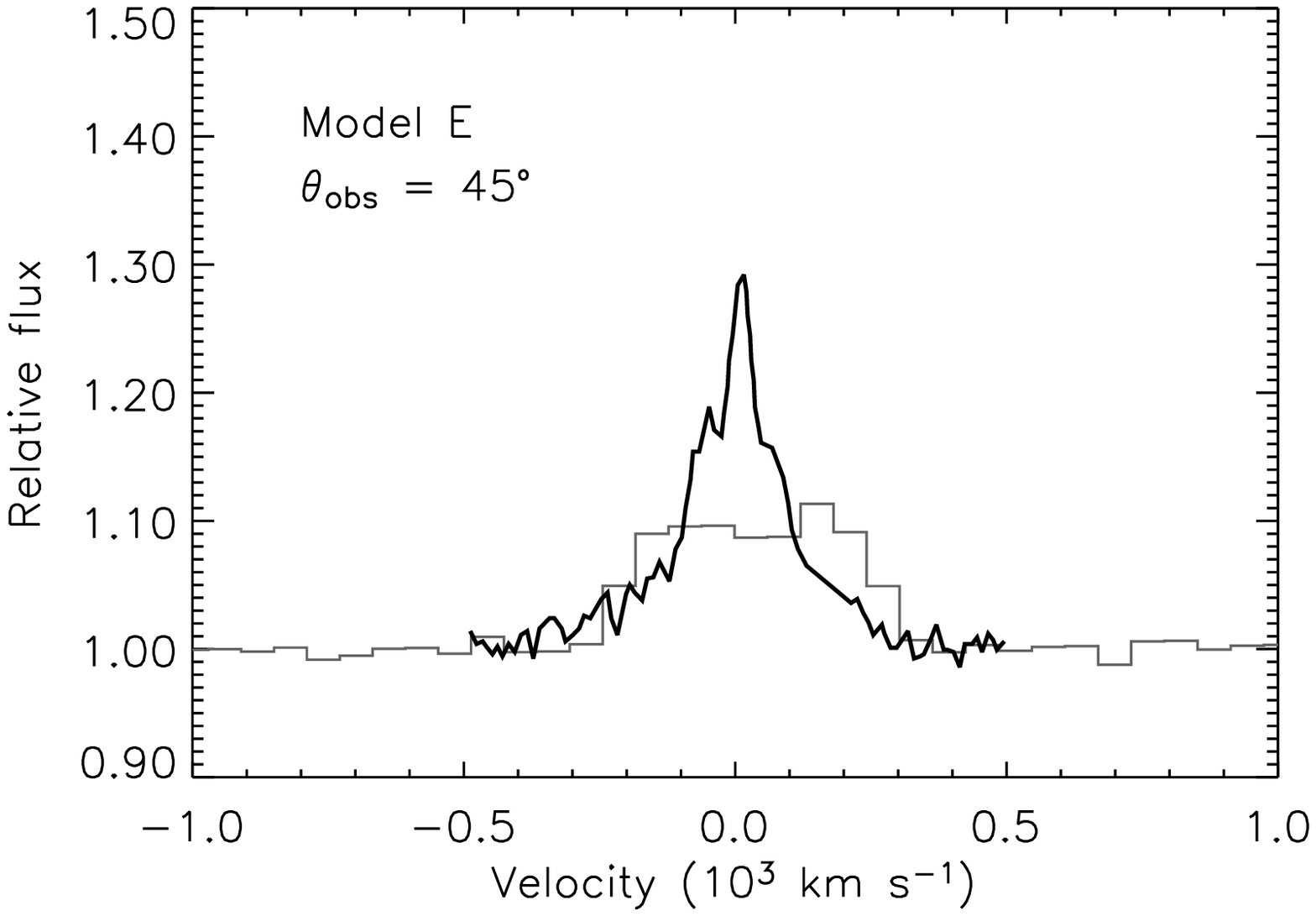, width=7cm}
\caption{Comparison of the observed Br$\alpha$ profile for GL989 (taken from Bunn et al. 1995; heavy line) with computed profiles from
Model~B (for viewing angle of 60\textdegree; light histogram on left) and Model~E 
(for viewing angle of 45\textdegree; light histogram on right). These models and angles
have been chosen for comparison of profile shapes since they give Br$\alpha$ equivalent-widths which are similar to
that observed for GL989.
We show Br$\alpha$ only -- since there are discrepancies between the modelled and
observed EWRs (see text), the models shown 
significantly underpredict the Br$\gamma$-EW and therefore do not simultaneously allow useful
comparisons of line shapes for all lines. Comparing the observed Br$\gamma$ or Pf$\gamma$ profiles to models
which reproduce their EWs leads to the same conclusions obtained from considering Br$\alpha$ only.}
\end{figure*}

Important differences are also evident in comparisons of the 
line profiles obtained from our models and those typically observed in massive YSOs.
To illustrate this, Figure~10 compares the observed Br$\alpha$ profile for GL989 (taken from Bunn et al. 1995)
with computed profiles from two of the models. GL989 is chosen here since the line profiles
for this object have strong extended wings and, in contrast to
several of the other objects in the Bunn et al. (1995) sample (e.g. S106IR), 
have simple profiles suggesting a single dominant region of line formation.
The comparison models and viewing angles
 (Model~B with $\theta_{\mbox{\scriptsize obs}} = 60$\textdegree~
and Model~E with $\theta_{\mbox{\scriptsize obs}} = 45$\textdegree) were selected since they 
predict Br$\alpha$ EWs close to the observed value, thereby allowing a direct comparison of profile shape. 

The figure shows that the Model~B Br$\alpha$ profile at $\theta_{\mbox{\scriptsize obs}} = 60$\textdegree~is  
too broad and, in contrast to the observations, is double-peaked. 
The Model~E profile ($\theta_{\mbox{\scriptsize obs}} = 45$\textdegree) is in better
agreement with the observation -- the combination of higher optical depth in the model (owing to the increased
mass-loss rate) and lower viewing angle reduce the profile width and suppress the double-peaks -- but a significant
discrepancy remains.
To resolve this difference in line shape between the models and typical 
observations, one could invoke even smaller inclination angles. 
Although viewing angles are not known for most 
luminous YSOs,
this is not an attractive solution since  
small viewing angles (close to pole-on) are statistically disfavoured 
in an unbiased sample. Furthermore, 
at least one of the objects discussed by Bunn et al. (1995) is known to be viewed almost edge-on
(S106IR; Solf \& Carsenty 1982).
Thus we prefer to pursue resolutions to the two styles of line shape 
discrepancies which do not rely on orientation effects.

Increasing the physical extent of the disk from which the
wind is launched does not readily eliminate the double-peaked line shape from our models (Models~D and E)
-- although some of the wind does occupy regions with lower rotational velocities when the disk is made larger, 
our assumption that the mass-loading is proportional to the luminous flux of the disk means that most of the mass in the
wind is still launched at small radii.
The problem of reconciling observed single-peaked line profiles with theoretical predictions of double peaks has
arisen elsewhere, particularly in studies of nova-like variables (see e.g. Horne 1997)
and
active galactic nuclei (AGN). In the context of AGN, Murray \& Chiang (1997) have argued that
substantial radial velocity shear in a disk wind can suppress the formation of double-peaked profiles as required.
For this to occur, it is necessary that line optical depths are high such that photon escape is strongly favoured in 
the poloidal direction where accelerating outflow ensures steep velocity gradients.  It is this difference between
Model~C, with enhanced mass loss, and Model~A that explains the relatively flat-topped profiles in Fig.~7, compared to 
Fig.~5.  Hence there is scope to re-organise model parameters in order to exploit this mechanism to suppress double
peaks.  
In particular, a way of raising line opacity without also raising Br$\alpha$ EWs is needed.
This will warrant further investigation in future work, focusing on fitting observations of specific objects.

Some of the discrepancies, in particular the large linewidths predicted by the models,
may be explained by departures from the particular disk wind geometry adopted (which was motivated
by the Drew et al. 1998 model).
Two interesting possibilities are those of an enlarged radius for the central star, relative to a main-sequence
radius, or of an inner gap in the YSO disk.
These 
possibilities have rather similar consequences --
by moving the disk wind out to larger radial distances
from the central object, the correspondingly lower 
rotation velocities reduce the extent of the 
line wings and make the double-peaks less prominent (this is illustrated by our model with a cavity in the inner disk, Model~F). 
Also, the SED for a disk
with a hole will be redder than that for a complete disk, helping to explain the EWR[Br$\alpha$/Br$\gamma$]
discrepancy. These alternatives might be distinguished by observational techniques that probe the inner
disk geometry (e.g. interferometry and spectropolarimetry) while the influence of a gap on the SED needs to be 
investigated fully in subsequent work incorporating more 
sophisticated treatments of the disk SED than used here.
 
A third possibility is that of disk winds in which the streamlines diverge less at large distances.
In our models, the divergence of the streamlines is imposed by our adopted ``split dipole'' geometry. 
In a geometry with less divergence of the flow, the density would be higher in the outer parts
of the wind for a given mass-loss rate and velocity law. 
The increased emission measure at large radii would help to wash out the double-peaked profiles
which form when the emission is dominated by gas in rotation close to the central object. For appropriate viewing angles, 
line-of-sight optical depths would also be larger for a less divergent outflow. As has already been noted, higher optical
depths present a promising solution to part of the EWR[Br$\alpha$/Br$\gamma$]
discrepancy. Thus, in the context of studying particular objects in detail, considering a range of different
disk wind geometries is likely to prove fruitful in future work.

These issues aside, the major success of our radiative transfer simulations 
is their support of the hypothesis that disk winds can play a significant role
in creating the H~{\sc i} lines of massive YSOs. 
In particular, the departure from
one-dimensional wind models has alleviated the need to invoke the uncomfortably large 
mass-loss rates required in earlier spherical models.
Further investigations of the applicability of the disk wind model to particular objects are now 
warranted.

\section*{Acknowledgements}

SAS thanks L. Lucy for several useful discussions relating to this work
and L. Mendes for technical support in use of the Imperial College 
Astrophysics Beowulf cluster.

This work was undertaken while SAS was a PPARC supported PDRA at 
Imperial College London (PPA/G/S/2000/00032).

\section*{References}

Bik A., Thi W. F., 2004, A\&A, 427, L13\\
Blum R. D., Barbosa C. L., Damineli A., Conti P. S., Ridgway S., \\
\indent 2004, ApJ, 617, 1167\\
Bonnell I. A., Vine S. G., Bate M. R., 2004, MNRAS, 349, 735\\
Bunn J. C., Hoare M. G., Drew J. E., 1995, MNRAS, 272, 346\\
Carr J. S., 1989, ApJ, 324, 522\\
Carr J. S., Tokunaga A. T., Najita J., Shu F. H., Glassgold A. E.,\\
\indent 1993, ApJL, 411, L37\\
Castor J. I., Abbott D. C., Klein R. I., 1975, ApJ, 409, 429\\
Chandler C. J., Carlstrom J. E., Scoville N. Z., Dent W. R. F.,\\
\indent Geballe T. R., 1993, ApJ, 412, L71\\
Chandler C. J., Carlstrom J. E., Scoville N. Z., 1995, ApJ, 446, 793\\
Cunto W., Mendoza C., Ochsenbein F., Zeippen C. J., 1993, A\&A,\\
\indent 275, L5\\
Drew J. E., Bunn J. C., Hoare M. G., 1993, MNRAS, 265, 12\\
Drew J. E., Proga D., Stone J. M., 1998, MNRAS, 296, L6\\
Dullemond C. P., Dominik C., Natta A., 2001, ApJ, 560, 957\\
Hamann F., Simon M., 1986, ApJ, 311, 909\\
Henning Th., Pfau W., Altenhoff W. J., 1990, A\&A, 227, 542\\
H\"{o}flich P., Wehrse R., 1987, A\&A, 185, 107\\
Hollenbach D., Johnstone D., Lizano S., Shu F., 1994, ApJ, 428,\\
\indent 654\\
Horne K., 1997, in Wickramasinghe D. T., Bicknell G. V., Ferrario\\
\indent L., eds, IAU Colloquium 163, ASP Conference Series, Vol.\\
\indent 121, p.14\\ 
Knigge C., Woods J. A., Drew J. E., 1995, MNRAS, 273, 225\\
Kraus M., Kr\"{u}gel E., Thum C., Geballe T. R., 2000, A\&A, 362,\\
\indent 158\\
Long K. S., Knigge C., 2002,  ApJ, 579, 725\\
Lucy L. B., Abbott D. C., 1993, ApJ, 405, 738\\
Lucy L. B., 2002, A\&A, 384, 725\\
Lucy L. B., 2003, A\&A, 403, 261\\
Menzel D. H., Pekeris C. L., 1935, MNRAS, 96, 77\\
Mihalas D., 1978, Stellar Atmospheres (2nd Edition; San\\
\indent Francisco: Freeman)\\
Monnier J. D. et al., 2005, astro-ph/0502252\\
Murray N., Chiang J., 1997, ApJ, 474, 91\\
Natta A., Prusti T., Neri R., Wooden D., Grimm V. P., Mannings\\
\indent V., 2001, A\&A, 371, 186\\
Proga D., Stone J. M., Drew J. E., 1998, MNRAS, 295, 595\\ 
Proga D., Stone J. M., Drew J. E., 1999, MNRAS, 310, 476\\ 
Simon M., Righini-Cohen G., Fischer J., Cassar L., 1981, ApJ,\\
\indent 251, 552\\
Simon M., Felli M., Cassar L., Fischer J., Massi M., 1983, ApJ,\\
\indent 266, 623\\ 
Solf J., Carsenty U., 1982, A\&A, 113, 142\\
Tuthill P. G., Monnier J. D., Danchi W. C., 2001, Nature, 409,\\
\indent 1012\\
van Regemorter H., 1962, ApJ, 136, 906\\
Vink J. S., Harries T. J., Drew J. E., 2005, A\&A, 430, 213\\

\label{lastpage}

\end{document}